\documentclass[11pt]{article}
\begin{document}

\newfam\msbfam
\batchmode\font\twelvemsb=msbm10 scaled\magstep1 \errorstopmode
\ifx\twelvemsb\nullfont\def\Bbb{\bf}
        \font\fourteenbbb=cmb10 at 14pt
	\font\eightbbb=cmb10 at 8pt
	\message{Blackboard bold not available. Replacing with boldface.}
\else   \catcode`\@=11
        \font\tenmsb=msbm10 \font\sevenmsb=msbm7 \font\fivemsb=msbm5
        \textfont\msbfam=\twelvemsb
        \scriptfont\msbfam=\tenmsb \scriptscriptfont\msbfam=\sevenmsb
        \def\Bbb{\relax\expandafter\Bbb@}
        \def\Bbb@#1{{\Bbb@@{#1}}}
        \def\Bbb@@#1{\fam\msbfam\relax#1}
        \catcode`\@=\active
	\font\fourteenbbb=msbm10 at 14pt
	\font\eightbbb=msbm8
\fi
\catcode`\@=11
\def\Z {{\Bbb Z}}
\def\R {{\Bbb R}}
\def\E {{\Bbb E}}
\newfam\scrfam
\batchmode\font\twelvescr=rsfs10 at 12pt \errorstopmode
\ifx\twelvescr\nullfont
        \message{rsfs script font not available. Replacing with calligraphic.}
        \def\scr{\cal}
\else   \font\tenscr=rsfs10 
        \font\sevenscr=rsfs7
        \skewchar\twelvescr='177 \skewchar\tenscr='177 \skewchar\sevenscr='177
        \textfont\scrfam=\twelvescr \scriptfont\scrfam=\tenscr
        \scriptscriptfont\scrfam=\sevenscr
        \def\scr{\fam\scrfam}
        \def\cal{\scr}
\fi
\def\unit{\hbox to 3.3pt{\hskip1.3pt \vrule height 7pt width .4pt \hskip.7pt
\vrule height 7.85pt width .4pt \kern-2.4pt
\hrulefill \kern-3pt
\raise 4pt\hbox{\char'40}}}
\def\II{{\unit}}
\def\cM {{\cal{M}}}
\def\half{{\textstyle {1 \over 2}}}
\newcommand{\od}{\widetilde{\rm OD}}
\def    \beq    {\begin{equation}} \def \eeq    {\end{equation}}
\def    \bea    {\begin{eqnarray}} \def \eea    {\end{eqnarray}}
\def\la{\label} \newcommand{\eq}[1]{(ref{#1})}
\def    \lf     {\left (} \def  \rt     {\right )}
\def    \a      {\alpha} \def   \lm     {\lambda}
\def    \D      {\Delta} \def   \r      {\rho}
\def    \th     {\theta} \def   \rg     {\sqrt{g}} \def \Slash  {\, /
\! \! \! \!}  \def      \comma  {\; , \; \;} \def       \pl
{\partial} \def         \del    {\nabla}
\newcommand{\mx}[4]{\left#1\begin{array}{#2}#3\end{array}\right#4}
\newcommand{\Dpp}{\Delta + \nu}
\newcommand{\Dmm}{\Delta - \nu}
\newcommand{\Dp}{\Delta_+}
\newcommand{\Dm}{\Delta_-}
\newcommand{\Ds}{\left(\Delta^2-\nu^2\right)}
\newcommand{\Pp}{\Pi_+}
\newcommand{\Pm}{\Pi_-}
\newcommand{\lp}{\ell_p}
\newcommand{\ie}{{\em i.e., }}
\newcommand{\eg}{{\em e.g., }}
\newcommand\sss{\scriptscriptstyle}
\newcommand\scs{\scriptstyle}
\newcommand{\bc}{\begin{center}}
\newcommand{\ec}{\end{center}}
\newcommand{\nz}{\normalsize}
\newcommand\nn{\nonumber}
\frenchspacing
\begin{titlepage}
\begin{flushleft}
       \hfill                      {\tt hep-th/0212107}\\
       \hfill                        G\"oteborg-ITP-preprint\\
\end{flushleft}
\vspace*{1.8mm}
\begin{center}
\LARGE {\bf An M-theory solution generating technique\\
and SL$(2,\mathbf{R})$}\\
\normalsize
\vspace*{5mm}
{\Large
Henric Larsson}\\
\vspace*{2mm}
{\small Institute of Theoretical  Physics\\
G\"{o}teborg University and Chalmers University of Technology\\
SE-412 96 G\"{o}teborg, Sweden \\
E-mail: solo@fy.chalmers.se}\\
\vspace*{3mm}
\end{center}
\Large \begin{center}
{\bf Abstract}
\end{center}
\normalsize
In this paper we generalize the $O(p+1,p+1)$ solution generating technique 
(this is a method used to deform D$p$-branes by turning on a NS-NS $B$-field) 
to M-theory, in order to be able to deform M5-brane supergravity solutions 
 directly in eleven dimensions, by turning on a non zero three form $A$. 
We find that deforming the M5-brane, in some cases, corresponds to performing 
certain SL$(2,\mathbf{R})$ transformations of the K$\ddot{{\rm a}}$hler 
structure 
parameter for the three-torus, on which the M5-brane has been compactified.
We show that this new M-theory solution generating technique can be reduced to 
the $O(p+1,p+1)$ solution generating technique with $p=4$. Further, we find 
that it 
implies that the open membrane metric and generalized noncommutativity 
parameter are manifestly deformation independent for electric and light-like 
deformations. We also generalize the $O(p+1,p+1)$ method to the type IIA/B 
NS5-brane in order to be able to deform NS5-branes with RR three and two 
forms, respectively. In the type IIA case we use the newly obtained solution 
generating technique and deformation independence to derive a covariant 
expression for an open D2-brane coupling, relevant for OD2-theory.

\end{titlepage}
\section{Introduction}
Recently there has been a lot of interest in theories with noncommutativity 
(see e.g., \cite{con}-\cite{davidper}). Using the AdS/CFT correspondence these
 theories can be studied using supergravity duals (see e.g., 
\cite{moto,maldaR,oz,gopa,Berman,mikkel,NS5,Delle,solo1}). These supergravity 
duals are obtained by taking a near horizon limit of a supergravity solution 
corresponding to some bound state. To obtain the relevant supergravity 
solution (bound state), one can either solve the equations of motion 
\cite{martin1,martin}, or start with a known brane solution and use some 
kind of 
solution generating technique \cite{moto,maldaR,Delle,Berman}. One solution 
generating technique which has been very useful is the so called 
$O(p+1,p+1)$ method \cite{Delle,Berman,solo1, solo2}, which uses elements of 
the T-duality group to generate bound states (not including the NS5-brane) in 
type IIA/B supergravity. This method can be seen as deforming D-branes by 
turning on an NS-NS two form $B$.
An important consequence of this solution generating technique is that 
it leaves the open string metric and coupling constant invariant and shifts 
the noncommutativity parameter by a constant \cite{Berman} (see also 
\cite{int}). This is referred to as
deformation independence of open string data. In \cite{openm} the concept of 
deformation independence was used to derive the open membrane metric and 
generalized noncommutativity parameter in eleven dimensions, see also 
\cite{janpieter,ericjanpieter}. 

In this paper we are going to generalize the $O(p+1,p+1)$ method to M-theory
 in order to obtain a formula for deforming M5-branes with a three form 
directly in eleven dimensions. We also generalize the $O(p+1,p+1)$ method in 
order to deform the NS5-brane in type IIA/B
with a RR three form and RR two form, respectively. These solution 
generating techniques can not be obtained as easily as the $O(p+1,p+1)$ method 
since they must include U-duality transformations and not only T-duality 
transformations. Here, we will obtain them in a more indirect way and check 
that 
they are consistent, e.g., we show that the eleven-dimensional supergravity 
`tensor' equation of motion is satisfied and that the M5-brane method when 
reduced to ten dimensions gives the $O(p+1,p+1)$ method for $p=4$. However, in
 the type IIB NS5-brane case we show how the deformed (D1) metric and RR two 
form can be derived using certain projective transformations of a tensor 
$F_{\mu\nu}$ built from the metric and RR two form, similar to how the metric 
and NS-NS two form are derived when deforming D$p$-branes with the 
$O(p+1,p+1)$ method. 

The main reason we derive these solution generating techniques is not in order
 to be able to obtain new bound states, but rather to obtain more 
information about symmetries of M-theory, U-duality and how the metric and
 three form transform under deformations of the M5-brane. In this paper we find
 that deforming M5-branes in some cases correspond to certain 
SL$(2,\mathbf{R})$ transformations of the complex scalar $E$, which is a 
certain combination of the determinant of the metric and the three 
form\footnote{The complex scalar $E$ is the K$\ddot{{\rm a}}$hler 
structure parameter for a three-torus, which the M5-brane has been 
compactified on. For related results we refer to \cite{sen1,aha1}.}. 
Unfortunately we do not obtain a \emph{complete} understanding of how the 
metric and three form transform under a deformation of the M5-brane. This will
 be discussed further in a future paper \cite{solo3}. A more complete 
understanding would be important to obtain since this might, e.g., give 
further important information
 about the open membrane metric, which is invariant under these kinds of 
transformation (in the electric and light-like cases). However, it is 
possible that the results obtained in this paper can give some hints. 

In section 2 we give a short introduction to the $O(p+1,p+1)$ method, which 
is of relevance for the rest of the paper. This is followed in section 3.1
 by a generalization of the $O(p+1,p+1)$ method to M-theory. In the rest of 
section 3 we perform several consistency checks in eleven dimensions. Next, in
 section 4 we 
reduce the M-theory method to ten dimensions and show that it gives the 
$O(p+1,p+1)$ method for $p=4$ and rank 2 NS-NS $B$-field. We further show that 
reducing the M-theory method transverse to the deformation directions, leads 
to a method for deforming a type IIA D4-brane with a RR three form.  
In section 5 we obtain a solution generating technique for deforming a type 
IIA/B NS5-branes with a RR three or two form, respectively. We also show 
that the two cases are T-dual to each other. Further, for the type IIA case, 
using the newly obtained solution generating 
technique and deformation independence we derive a covariant expression 
for an open D2-brane coupling, relevant for OD2-theory \cite{GMSS}. Next, in
section 6 we show that in certain cases deforming the M5/NS5-branes 
involves certain SL$(2,\mathbf{R})$ transformations of the 
K$\ddot{{\rm a}}$hler structure parameter $E$. We end with some 
conclusions in section 7.

\section{The $O(p+1,p+1)$ method}
In this section we give a short review of the $O(p+1,p+1)$ solution generating
 technique. For more details see \cite{Delle}. 

For a NS-NS $B$-field deformation of a general D$p$--brane one first 
T-dualizes 
in the directions where one wants to turn on NS-NS fluxes, and then one
shifts the $B$-field with a constant in these directions. After this
one T-dualizes in the directions where one has turned on the constant 
$B$-field. In a more precise language, the
deformation with constant parameter $\theta_{\rm s}^{\mu\nu}$ is generated by
the following $O(p+1,p+1)$ T-duality group element\footnote{See 
\cite{Delle,Berman} for conventions and definitions of the 
various elements of $O(p+1,p+1)$ appearing in the following discussion.}
\beq \label{r}
\Lambda=\Lambda_0\dots
\Lambda_p\Lambda_{-\theta_{\rm s}}\Lambda_p\cdots \Lambda_0=
J\Lambda_{-\theta_{\rm s}}J=\Lambda_{\theta_{\rm s}}^T =
\mx{(}{ll}{1&0\\ \theta_{\rm s}&1}{)}\ ,  \eeq
where $\theta_{\rm s}^{\mu\nu}$ is dimensionless and carries indices
upstairs since it starts life on the T-dual world volume \cite{Delle}. In 
(\ref{r}) above, $\Lambda_{i}$ ($i=0,\ldots ,p$) corresponds to a T-duality 
transformation in the $i$:th direction, while $\Lambda_{-\theta_{{\rm s}}}$ 
corresponds 
to a constant shift in $B_{2}$ (i.e., a gauge transformation).

Starting with a D$p$--brane solution\footnote{We are using
 multi-form notation such that $C$ is a sum of forms while $B$ (see below)
 has fixed rank $2$.}
\begin{eqnarray}\label{ud}
ds^{2}&=&g^{\rm (s)}_{\mu\nu}dx^{\mu}dx^{\nu}+g^{\rm (s)}_{mn}dx^{m}dx^{n}\ ,
\quad e^{2\phi}=g^{2}F\ , 
\nonumber\\
gC&=&\omega dx^{0}\wedge \cdots \wedge dx^{p}+\gamma_{7-p}\ ,
\end{eqnarray}
where $F$ is some function, $\omega=ge^{-\phi}\sqrt{-g^{\rm (s)}_{(\mu\nu)}}$, 
 ($g^{\rm (s)}_{(\mu\nu)}={\rm det}g^{\rm (s)}_{\mu\nu}$) due to the zero 
force condition, $g$ is the closed string coupling constant and 
$x^{\mu}$, $\mu=0,\ldots ,p$, are coordinates in the brane directions,  
while $x^{m}$, $m=p+1,\ldots ,9$, are coordinates in the transverse 
directions.
 Also, $\gamma_{7-p}$ is a transverse form, i.e., $i_{\mu}\gamma_{7-p}=0$, 
where $i_{\mu}$ denotes the inner product with the vector field associated 
with $x^{\mu}$. We note that it is only possible to deform (i.e., turn on 
a non-zero NS-NS two form $B$) the D$p$-brane in those directions in which we 
can use T-duality. This is a constraint on (\ref{ud}).

For the NS-NS fields $g^{\rm (s)}_{\mu\nu}$ and $B_{\mu\nu}$, the 
transformations in (\ref{r}) imply that the tensor $E_{\mu\nu}=g^{\rm 
(s)}_{\mu\nu}+B_{\mu\nu}$ transforms by the following projective transformation
\cite{Delle,Berman} (Note that in (\ref{ud}) \mbox{$B_{\mu\nu}=0$})
\begin{equation}\label{r1}
\tilde{E}_{\mu\nu}=\Big(\frac{E}{\theta_{\rm s} E+1}\Big)_{\mu\nu}=\Big(\frac{
g^{\rm (s)}(1-\theta_{\rm s} g^{\rm (s)})}{(1+\theta_{\rm s} g^{\rm (s)})
(1-\theta_{\rm s} g^{\rm (s)}) }\Big)_{\mu\nu}\ .
\end{equation}
Now using (\ref{r1}) and how the dilaton and the RR fields transform (see 
\cite{Delle,Berman}) we obtain the following deformed D$p$-brane configuration:
\begin{eqnarray}\label{def}
\tilde{g}^{\rm (s)}_{\mu\nu}&=&g^{\rm (s)}_{\mu\rho}\Big[(1-(\theta_{\rm s})
^{2})^{-1}\Big]^{\rho}{}_{\nu}\ , 
\quad \tilde{g}^{\rm (s)}_{mn}=g^{\rm (s)}_{mn}\ , \nonumber\\
\tilde{B}_{\mu\nu}&=&-g^{\rm (s)}_{\mu\rho}\theta_{\rm s}^{\rho\sigma}
g^{\rm (s)}
_{\sigma\lambda}\Big[(1-(\theta_{\rm s})^{2})^{-1}\Big]^{\lambda}{}_{\nu}\ ,\\
e^{2\tilde{\phi}}&=&\frac{e^{2\phi}}{\sqrt{{\rm det}(1-(\theta_{\rm s})^{2})}}=
e^{2\phi}\Big(\frac{{\rm det}\tilde{g}}{{\rm det}g}\Big)^{\frac{1}{2}}\ ,
\nonumber\\
g\tilde{C}&=&e^{-\frac{1}{2}\tilde{B}_{\mu\nu}dx^{\mu}\wedge dx^{\nu}}\big(
\omega e^{\frac{1}{2}\theta_{\rm s}^{\mu\nu}i_{\mu}i_{\nu}}dx^{0}\wedge\cdots 
\wedge dx^{p}+\gamma_{7-p}\big)\ ,\nonumber
\end{eqnarray}
where
\begin{equation}
(\theta_{\rm s}^{2})^{\mu}{}_{\nu}=\theta_{\rm s}^
{\mu\nu'}g^{\rm (s)}_{\nu'\rho}\theta_{\rm s}^{\rho\sigma}
g^{\rm (s)}_{\sigma\nu}\ .
\end{equation}

There are two types of deformations that are possible: $\theta^{0i}$ and 
$\theta^{ij}$, where $i,j=1,2,\ldots,p$. The first one is called `electric' 
since we mix the time direction with a spatial direction,
while the second is called `magnetic' since the time direction is not included.
Further, we note that the deformed solution (\ref{def}) satisfies the zero 
force condition, because the zero force condition is satisfied by the 
undeformed solution (\ref{ud}), see \cite{Delle,Berman}. The deformed solution 
also preserves the same amount of supersymmetry as the undeformed solution.

Before we end this review section we note that there is a very simple 
relation between the open string metric, noncommutativity parameter and 
the closed string metric and NS-NS $B$-field. Defining 
\begin{equation}
\tau^{\mu\nu}=G_{(\rm s)}^{\mu\nu}+\frac{\Theta^{\mu\nu}}{\alpha'}\ ,
\end{equation}
it is easy to obtain that $E$ and $\tau$ are related through
\begin{equation}\label{relos}
\tau^{\mu\nu}=(E^{-1})^{\mu\nu}\ .
\end{equation}
In later sections we will see that similar relations hold also for open 
D-brane data and closed D-brane data, see in particular section 5. 

\section{An M-theory solution generating technique and some tests}
In this section we will argue for the existence of an M-theory solution 
generating technique that can be used to deform M5-branes with a non-zero 
three form $A$, which obeys a non-linear self-duality equation. As we will see 
below, this is (as expected) the same non-linear self-duality 
equation as the gauge invariant M5-brane world volume three form $\mathcal{H}$
 satisfies. 

In section 3.1 
we \emph{conjecture} the exact form of this solution generating technique, 
while in the following subsections we test the conjecture in eleven dimensions.
\subsection{An M-theory solution generating technique}
Here we will generalize the $O(p+1,p+1)$ method to eleven dimensions, 
in order to deform M5-branes with a non-zero three form $A$. We start with the
 following (general) M5-brane solution\footnote{Note that this M5-brane 
solution is assumed to reduce to (\ref{ud}), with $p=4$.}:
\begin{eqnarray}\label{M5g}
ds^{2}&=&g_{\mu\nu}dx^{\mu}dx^{\nu}+g_{mn}dx^{m}dx^{n}\ ,\quad \mu,\nu=0,1,
\ldots ,5\ ,\quad m,n=6,7,8,9,10\ ,
\nonumber\\
A_{6}&=&\omega dx^{0}\wedge \cdots \wedge dx^{5}\ ,\quad A_{3}=\gamma_{3}\ ,
\end{eqnarray} 
where $\gamma_{3}$ is a transverse three form dual to the six form (i.e., 
$\ast d\gamma_{3}=dA_{6}$), while 
$\omega=\sqrt{-g_{(\mu\nu)}}$ due to the zero force condition and the metric 
is assumed to be diagonal. 

An important difference between the $O(p+1,p+1)$ 
method and a method for deforming M5-branes is that the former can be derived 
from the T-duality group, while the latter does not seem to be possible to 
derive because we lack a microscopic formulation of M-theory. We will 
therefore conjecture the exact form of the M-theory solution generating 
technique and test this conjecture both directly in eleven dimensions and 
show that it reduces to known results in ten dimensions. We now conjecture 
that the generalization of (\ref{def}) to deformations of the 
eleven-dimensional M5-brane with a non-zero three form $A$, is given by
\begin{eqnarray}\label{defM}
\tilde{g}_{\mu\nu}&=&\Big[{\rm det}\Big(1+\frac{1}{2}(\theta)^{2}\Big)
\Big]^{1/9}g_{\mu\rho}\Big[\Big(1+\frac{1}{2}(\theta)^{2}\Big)^{-1}
\Big]^{\rho}{}_{\nu}\ , \quad 
\tilde{g}_{mn}=\Big[{\rm det}\Big(1+\frac{1}{2}(\theta)^{2}\Big)\Big]^{1/9}
g_{mn}\ , \nonumber\\
\tilde{A}_{3}&=&\tilde{A}_{3{\rm a}}+\tilde{A}_{3{\rm b}}+\gamma_{3}\ ,
\nonumber\\
\tilde{A}_{6}&=&A_{6}+\frac{1}{2}\tilde{A}_{3{\rm a}}\wedge 
\tilde{A}_{3{\rm b}}+\frac{1}{2}(\tilde{A}_{3{\rm a}}+ \tilde{A}_{3{\rm b}})
\wedge \gamma_{3}\ ,\\
{\rm where}\quad & &\nonumber\\
\tilde{A}_{3{\rm a}}&=&\frac{1}{6}\tilde{A}^{3{\rm a}}_{\mu\nu\rho}
dx^{\mu}\wedge  dx^{\nu}\wedge dx^{\rho}\ ,\quad 
\tilde{A}^{3{\rm a}}_{\mu\nu\rho}=-g_{\mu\rho'}g_{\sigma\nu}
\theta^{\rho'\sigma\sigma'}g_{\sigma'\lambda}\Big[\Big(1+\frac{1}{2}
(\theta)^{2}\Big)^{-1}\Big]^{\lambda}{}_{\rho}\ ,\nonumber \\
\tilde{A}_{3{\rm b}}&=&-\frac{1}{6}\omega \theta^{\mu\nu\rho}i_{\mu}i_{\nu}
i_{\rho}dx^{0}\wedge \cdots \wedge dx^{5}\ .\nonumber 
\end{eqnarray}
Here $\theta^{\mu\nu\rho}$ is a constant dimensionless anti-symmetric 
 tensor, and $(\theta^{2})^{\mu}{}_{\nu}$ is defined as follows:
\begin{equation}
(\theta^{2})^{\mu}{}_{\nu}=\theta^{\mu\nu'\rho}g_{\nu'\sigma}g_{\rho\sigma'}
\theta^{\sigma\sigma'\lambda}g_{\lambda\nu}\ .
\end{equation}

The deformation parameter $\theta^{\mu\nu\rho}$ is constrained 
to only have `one' non-zero component, e.g., 
$\theta^{\alpha\beta\gamma}=\theta\epsilon^{\alpha\beta\gamma}$ 
($\epsilon^{012}=1$), where $\alpha=0,1,2,$ while 
$\theta^{abc}=0$, $a=3,4,5.$ It has been shown in \cite{martin,us2,openm} 
that one parameter is enough to parameterize all deformations of an M5-brane 
with a non-linearly self-dual three form $A$ 
(up to Lorentz transformations). It is therefore no restriction to constrain 
$\theta^{\mu\nu\rho}$ to only have one parameter. An important difference 
between the M5-brane and D4-brane is that from an M5-brane point of view 
the difference between rank 2 and rank 4 $B$-field on the D4-brane is a 
Lorentz transformation (see e.g., \cite{ericjanpieter}).

This solution generating technique works similarly to the $O(p+1,p+1)$ method. 
This means that if we, e.g., want to turn on a non-zero magnetic three form, 
we start with an undeformed M5-brane which we compactify on a three-torus. 
Next, we invert the volume of the torus, which implies that 
the M5-brane becomes an M2-brane smeared in three directions. This is 
followed by a gauge transformation of the three form $A$ in the three 
directions which the M2-brane is smeared, e.g., $A_{345}=0\rightarrow 
A_{345}=\theta$. Finally, we invert the volume of the three-torus and then we  
decompactify. This interpretation of the solution generating technique 
will be further motivated in section 6.

Next, we are going to give the following arguments why (\ref{defM}) is 
correct. 

\begin{enumerate} 

\item{We show that a double dimensional reduction along the $y$ direction to a
 rank 2 $\theta^{\mu\nu}_{\rm s}=\theta^{\mu\nu y}$, gives (\ref{def}) 
constrained to rank 2.}

\item{We will show that (\ref{defM}) generates the correct half-supersymmetric
 M5-M2 and M5-M2-M2-MW bound states. This further implies that (\ref{defM}) 
gives the correct solution if we start with an undeformed M5-brane solution 
with a conformally flat metric in the M5-brane directions, see section 3.2.}

\item{We will also show that (\ref{defM}) implies that the open membrane 
metric is manifestly deformation independent under  
electric or light-like deformations, while the generalized noncommutativity 
parameter is constant as expected. We note that since
an electric and a magnetic deformation must be related through a coordinate 
transformation, the open membrane metric is obviously \emph{not} deformation 
independent under a magnetic deformation. However, the generalized 
noncommutativity parameter is still constant. The reason that deformation 
independence is important to show is because it was used in the construction 
of the open membrane metric in \cite{openm}. However, there deformation 
independence was \emph{assumed} for \emph{one} particular deformed solution. 
Here we show that any electric (and light-like) deformation gives manifestly 
deformation independent solutions, see section 3.3.}

\item{Further, in section 3.4 we show that (\ref{defM}) satisfies the 
non-linear self-duality 
equation for the three form $A$ (in the M5-brane directions). We further show  
that $\ast \tilde{H}_{4}=\tilde{H}_{7}$ and most importantly that the 
eleven-dimensional supergravity `tensor' 
equation of motion for the three form $A$ is satisfied.}

\item{We are also going to show that a reduction to type IIA with  
 $\theta^{\mu\nu\rho}$, gives a new formula for one 
parameter RR three form deformations of D4-branes. This new formula is
 shown to give all the expected results, see section 4.2. For a relation to 
a formula for one parameter RR three form deformations of NS5-branes, see 
section 5.2.}

\item{Finally, we show that for a magnetic deformation, the deformed metric and
 three form (in the deformed directions) can be obtained by considering
 certain SL$(2,\mathbf{R})$ transformations of the K$\ddot{\rm a}$hler 
structure parameter for the three-torus, on which the M5-brane is 
compactified, see section 6.}
\end{enumerate}

Together, these tests strongly indicate that the solution generating 
technique (\ref{defM}) is correct. However, they are not enough to rigorously 
prove that (\ref{defM}) is correct. A rigorous proof would be to also show that
 the eleven-dimensional `Einsteins' equation of motion is satisfied by 
(\ref{defM}). This, however, seems to be very difficult to show, because 
`Einsteins' equation of motion is second order in derivatives of the deformed 
metric. This is an important difference (complication) compared to the 
`tensor' equation, which avoids explicit derivatives on the deformed metric, 
see section 3.4. A complete proof of (\ref{defM}) could also be obtained 
if there existed a microscopic formulation of M-theory `T-duality'. Then 
(\ref{defM}) could be derived from the M-theory `T-duality' rules. In a 
future paper \cite{solo3} we plan to further investigate the latter 
possibility.

\subsection{M5-M2 and M5-M2-M2-MW solutions} 

In this subsection we will show that using (\ref{defM}) gives the correct 
half-supersymmetric supergravity solutions corresponding to M5-M2 and 
M5-M2-M2-MW bound states. 

We begin by giving the half-supersymmetric M5-brane solution \cite{M5f}:
\begin{eqnarray}\label{M5}
ds^{2}&=&H^{-\frac{1}{3}}\eta_{\mu\nu}dx^{\mu}dx^{\nu}+
H^{\frac{2}{3}}\delta_{mn}dx^{m}dx^{n}\ ,\quad \mu,\nu=0,1,
\ldots ,5\ ,\quad m,n=6,7,8,9,10\ ,
\nonumber\\
A_{6}&=&H^{-1} dx^{0}\wedge \cdots \wedge dx^{5}\ ,\quad A_{3}=\gamma_{3}\ ,
\quad H=1+\frac{R^{3}}{r^{3}}\ ,
\end{eqnarray} 
where $H$ is a harmonic function on the transverse space and 
$\gamma_{3}$ is the three form dual to the six form, i.e., $\gamma=3R^{3}
\epsilon_{3}$, where $d\epsilon_{3}$ is the volume form of the four-sphere.
Now, continuing by deforming this M5-brane solution with 
$\theta^{012}=\theta$, using (\ref{defM}), gives\footnote{As far as we know 
this is the first time the M5-M2 brane solution has been given with both the 
three form and the dual six form. We note that the 
M5-M2 bound state was first obtained in \cite{iz}. Note that their solution is
 written in a different form than ours.}
\begin{eqnarray}\label{M5M2e}
d\tilde{s}^{2}&=&(Hh_{-}^{-1})^{-\frac{1}{3}}\Big[\frac{1}{h_{-}}
\Big((dx^{0})^{2}+
(dx^{1})^{2}+(dx^{2})^{2}\Big)+(dx^{3})^{2}+(dx^{4})^{2}+(dx^{5})^{2}\Big] 
\nonumber\\
& &+h_{-}^{\frac{1}{3}}H^{\frac{2}{3}}\delta_{mn}dx^{m}dx^{n}\ , \nonumber\\
\tilde{A}_{3}&=&\frac{\theta}{Hh_{-}}dx^{0}\wedge dx^{1}\wedge dx^{2}+
\frac{\theta}{H}
dx^{3}\wedge dx^{4}\wedge dx^{5}+\gamma_{3}\ ,\\
\tilde{A}_{6}&=&(Hh_{-})^{-1}\Big(\frac{h_{-}+1}{2}\Big)dx^{0}\wedge \cdots 
\wedge dx^{5}\nonumber\\
&+&\frac{1}{2}\theta H^{-1}(h_{-}^{-1}dx^{0}\wedge dx^{1}
\wedge dx^{2}+dx^{3}\wedge dx^{4}\wedge dx^{5})\wedge \gamma_{3}\ ,
\nonumber\\
h_{-}&=&1-\theta^{2}H^{-1}\ ,
\nonumber
\end{eqnarray} 
while a deformation with $\theta^{345}=-\theta'$, gives
\begin{eqnarray}\label{M5M2m}
d\tilde{s}^{2}&=&(Hh_{+}^{-1})^{-\frac{1}{3}}\Big[(dx^{0})^{2}+
(dx^{1})^{2}+(dx^{2})^{2}+\frac{1}{h_{+}}\Big((dx^{3})^{2}+(dx^{4})^{2}+
(dx^{5})^{2}\Big)\Big]\nonumber \\
& &+h_{+}^{\frac{1}{3}}H^{\frac{2}{3}}\delta_{mn}dx^{m}dx^{n}\ , \nonumber\\
\tilde{A}_{3}&=&\frac{\theta'}{H}dx^{0}\wedge dx^{1}\wedge dx^{2}+
\frac{\theta'}
{Hh_{+}}dx^{3}\wedge dx^{4}\wedge dx^{5}+\gamma_{3}\ ,\\
\tilde{A}_{6}&=&(Hh_{+})^{-1}\Big(\frac{h_{+}+1}{2}\Big)dx^{0}\wedge \cdots 
\wedge dx^{5}\nonumber\\
&+&\frac{1}{2}\theta' H^{-1}(dx^{0}\wedge dx^{1}\wedge dx^{2}+h_{+}^{-1}
dx^{3}\wedge dx^{4}\wedge dx^{5})\wedge \gamma_{3}\ ,\nonumber\\
h_{+}&=&1+\theta'^{2}H^{-1}\ .\nonumber
\end{eqnarray} 

Finally, deforming (\ref{M5}) with a light-like $\theta^{-12}=\theta$ 
(where $x^{\pm}=\frac{1}{\sqrt{2}}(x^{5}\pm x^{0})$), gives
\begin{eqnarray}\label{M5LL}
d\tilde{s}^{2}&=&H^{-\frac{1}{3}}(2dx^{-}dx^{+}-\theta^{2}H^{-1}(dx^{+})^{2}+
(dx^{1})^{2}+(dx^{2})^{2}+(dx^{3})^{2}+(dx^{4})^{2})\nonumber\\
& &+H^{\frac{2}{3}}\delta_{mn}dx^{m}dx^{n}\ ,\nonumber\\
\tilde{A}_{6}&=&H^{-1}dx^{0}\wedge \cdots \wedge dx^{5}\nonumber\\
&-&\frac{1}{2}\theta H^{-1}dx^{+}\wedge (dx^{1}\wedge dx^{2}+dx^{3}\wedge 
dx^{4})\wedge \gamma_{3}\ ,\\
\tilde{A}_{3}&=&-\frac{\theta}{H}dx^{+}\wedge (dx^{1}\wedge dx^{2}+
dx^{3}\wedge dx^{4})+\gamma_{3}\ .\nonumber
\end{eqnarray}
Note that if we instead deform with $\theta^{-34}=-\theta$, we obtain the 
same result as in (\ref{M5LL}). 

Of these three solutions, the first two correspond to equivalent M5-M2 
bound states\footnote{Further, after taking appropriate near horizon limits 
(electric and magnetic, respectively, see \cite{Berman}) of (\ref{M5M2e}) and 
(\ref{M5M2m}) we obtain solutions which are not only equivalent but 
identical. This solution is the supergravity dual of OM-theory 
\cite{GMSS,harmark2}.}, while the third corresponds to an M5-M2-M2-MW 
bound state with equal absolute value of the M2-brane charges.

\subsection{Open membrane data and deformation independence}
Here we show that (\ref{defM}) inserted in the open membrane metric and 
generalized noncommutativity (theta) parameter, gives the expected results, 
i.e., 
for an electric or a light-like deformation $\tilde{G}^{\sss 
{\rm OM}}_{\mu\nu}$ is independent of the deformation parameter $\theta$, 
while $\Theta_{\sss {\rm OM}}^{\mu\nu\rho}$ is constant. 
The open membrane metric and generalized noncommutativity parameter are given 
by \cite{janpieter,openm,ericjanpieter}\footnote{For related work concerning 
a three index structure and open membranes, see \cite{three}.}:
\begin{eqnarray}\label{OMdata}
\tilde{G}_{\mu\nu}^{{\sss {\rm OM}}}&=&\Big(\frac{1-\sqrt{1-K^{-2}}}{K^{2}}
\Big)^{1/3}\Big(\tilde{g}_{\mu\nu}+\frac{1}{4}(\tilde{A}^{2})_{\mu\nu}\Big)\ ,
\\ \Theta^{\mu\nu\rho}_{{\sss {\rm OM}}}&=&-\ell_{\rm p}^{3}
[K(1-\sqrt{1-K^{-2}})]^{2/3}\tilde{g}^{\mu\mu_{1}}
\tilde{A}_{\mu_{1}\nu_{1}\rho_{1}}\tilde{G}^{\nu_{1}\nu}_{{\sss {\rm OM}}}
\tilde{G}^{\rho_{1}\rho}_{{\sss {\rm OM}}}\ ,\nonumber
\end{eqnarray}
where
\begin{equation}
(\tilde{A}^2)_{\mu\nu}=\tilde{g}^{\mu_{1}\nu_{1}}\tilde{g}^{\mu_{2}\nu_{2}}
\tilde{A}_{\mu_{1}\mu_{2}\mu}\tilde{A}_{\nu_{1}\nu_{2}\nu}\ ,\quad 
\tilde{A}^{2}=\tilde{g}^{\mu\nu}(\tilde{A}^{2})_{\mu\nu}\ ,\quad 
K=\sqrt{1+{1\over 24}\tilde{A}^2}\ .
\end{equation}
To simplify the calculations (for the electric and magnetic cases) we use that
 if one goes to a frame 
$(u^{\alpha}_{\mu},v^{a}_{\mu})$, where $\alpha=0,1,2$ and $a=3,4,5$, 
parameterize the coset SO(1,5)/SO(1,2)$\times$SO(3) 
as defined in \cite{us2}, then the open membrane metric and theta parameter 
can be written as:
\begin{eqnarray}\label{ommetric}
\tilde{G}^{{\sss {\rm OM}}}_{\alpha\beta}&=&\Big(1+\frac{1}{6}
\tilde{A}_{1}^{2}\Big)^{2/3}\tilde{g}_{\alpha\beta}\ ,
\quad \tilde{A}_{1}^{2}=\tilde{g}^{\alpha\beta}(\tilde{A}_{1}^{2})_{\alpha
\beta}\ ,\nonumber\\ 
(\tilde{A}_{1}^{2})_{\alpha\beta}&=&\tilde{g}^{\alpha_{1}\beta_{1}}\tilde
{g}^{\alpha_{2}\beta_{2}}\tilde{A}_{\alpha_{1}\alpha_{2}\alpha}
\tilde{A}_{\beta_{1}\beta_{2}\beta}\ ,\quad \alpha,\beta=0,1,2\ ,\nonumber\\
\tilde{G}^{{\sss {\rm OM}}}_{ab}&=&\Big(1+\frac{1}{6}\tilde{A}_{2}^{2}
\Big)^{1/3}\tilde{g}_{ab}\ ,\quad \tilde{A}_{2}^{2}=\tilde{g}^{ab}(
\tilde{A}_{2}^{2})_{ab}\ ,\\
(\tilde{A}_{2}^{2})_{ab}&=&\tilde{g}^{a_{1}b_{1}}\tilde{g}^{a_{2}b_{2}}
\tilde{A}_{a_{1}a_{2}a}\tilde{A}_{b_{1}b_{2}b}\ ,\quad a,b=3,4,5\ ,\nonumber
\end{eqnarray}
and 
\begin{equation}\label{thetaabc}
\Theta^{\alpha\beta\gamma}_{\sss {\rm OM}}=-\ell_{\rm p}^{3}(1+
\frac{1}{6}\tilde{A}_2^2)\tilde{A}^{\alpha\beta\gamma}\ ,
\quad \Theta^{abc}_{\sss {\rm OM}}=-\ell_{\rm p}^{3}(1+
\frac{1}{6}\tilde{A}_1^2)\tilde{A}^{abc}\ .
\end{equation}
To obtain these relations we have used that
\begin{eqnarray}
K^{2}&=&\frac{(1+\frac{1}{12}\tilde{A}_{i}^{2})^{2}}{1+\frac{1}{6}\tilde
{A}_{i}^{2}}\ ,\quad i=1,2\ ,\nonumber\\
1+\frac{1}{6}\tilde{A}^{2}_{2}&=&\Big(1+\frac{1}{6}\tilde{A}^{2}_{1}
\Big)^{-1}\ ,
\end{eqnarray}
in the above parameterization. Note that equation (\ref{thetaabc}) implies 
that $\Theta^{\mu\nu\rho}_{\sss {\rm OM}}$ is completely anti-symmetric.
In \cite{openm} it is further shown, using (\ref{thetaabc}) and 
(\ref{ommetric}), that
\begin{equation}\label{duality}
\ast_{\tilde{G}} \Theta_{\sss {\rm OM}} = \Theta_{\sss {\rm OM}}\ ,\quad 
(\ast_{\tilde{G}} \Theta_{\sss {\rm OM}})^{\mu\nu\rho}=\frac{1}{6}\frac{1}
{\sqrt{-\tilde{G}}}\epsilon^{\mu\nu\rho\iota\kappa\lambda}
\Theta^{\sss {\rm OM}}_{\iota\kappa\lambda}\ ,
\end{equation}
where $\tilde{G}$ is the determinant of the open membrane metric and the 
indices on $\Theta_{\sss {\rm OM}}$ are lowered with 
$\tilde{G}^{{\sss {\rm OM}}}_{\mu\nu}$. From this relation we see that 
$\Theta^{\mu\nu\rho}_{\sss {\rm OM}}$ is \emph{linearly} self-dual with 
respect to the open membrane metric $\tilde{G}^{{\sss {\rm OM}}}_{\mu\nu}$.  

Next, we continue by simplifying (\ref{defM}) in the above parameterization. 
For an electric deformation, i.e., $\theta^{012}=\theta$, the metric and 
three form in (\ref{defM}), in the M5-brane directions, can be written as 
\begin{eqnarray}\label{simpe}
\tilde{g}_{\alpha\beta}&=&\Big(1+\frac{1}{6}(\theta)^{2}\Big)^{-2/3}
g_{\alpha\beta}\ ,
\quad \tilde{g}_{ab}=\Big(1+\frac{1}{6}(\theta)^{2}\Big)^{1/3}g_{ab}\ ,
\nonumber\\
\tilde{A}_{012}&=&-\theta g_{(\alpha\beta)}\Big(1+\frac{1}{6}(\theta)^{2}\Big)^
{-1}\ ,\quad \tilde{A}_{345}=\theta\omega=\theta\sqrt{-g_{(\mu\nu)}}\ ,
\end{eqnarray}
where $(\theta)^{2}=g^{\alpha\beta}(\theta^{2})_{\alpha\beta}$, 
$g_{(\alpha\beta)}$ is the determinant of $g_{\alpha\beta}$ 
and we have used that $(\theta^{2})_{\alpha\beta}=\frac{1}{3}g_{\alpha\beta}
(\theta)^{2}$, while $A_{1}^{2}=(\theta)^{2}$.  
For the purpose of later sections we note that 
\begin{equation}\label{121}
1+\frac{1}{6}(\theta)^{2}=1+\theta^{2}g_{(\alpha\beta)}\ .
\end{equation}
Inserting (\ref{simpe}) in the open membrane metric and theta parameter gives
\begin{equation}\label{simpe1}
\tilde{G}_{\mu\nu}^{\sss {\rm OM}}=g_{\mu\nu}\ , \quad \Theta^{012}_{\sss {\rm
 OM}}=\ell_{\rm p}^{3}\theta\ ,\quad \Theta^{345}_{\sss {\rm OM}}=-
\ell_{\rm p}^{3}\theta\sqrt{\frac{-g_{(\alpha\beta)}}{g_{(ab)}}}\ .
\end{equation}
We see here that the open membrane metric is deformation independent and 
in the case when $g_{\mu\nu}$ is conformally flat, the theta parameter is
$\Theta^{012}_{\sss {\rm OM}}=\Theta^{345}_{\sss {\rm OM}}=\ell_{\rm
 p}^{3}\theta$, which is the result we expected.  

Next, if we repeat the above analysis for a magnetic deformation 
$\theta^{345}=-\theta$, we obtain the following result:
\begin{equation}\label{simpm}
\tilde{G}_{\mu\nu}^{\sss {\rm OM}}=\Big(1+\frac{1}{6}(\theta)^{2}\Big)^{-1/3}
g_{\mu\nu}\ , \quad \Theta^{345}_{\sss {\rm OM}}=-
\ell_{\rm p}^{3}\theta\ ,\quad \Theta^{012}_{\sss {\rm OM}}=
\ell_{\rm p}^{3}\theta\sqrt{\frac{g_{(ab)}}{-g_{(\alpha\beta)}}}\ .
\end{equation}
Here the open membrane metric is not deformation independent, 
which we also expected, see \cite{openm}.
Further, since electric and magnetic deformations give equivalent M5-M2 bound 
states, it is clear that 
$\sqrt{\frac{g_{(ab)}}{-g_{(\alpha\beta)}}}={}$constant, in both 
(\ref{simpe1}) 
and (\ref{simpm}). This in turn implies that the theta parameter is 
always constant. 

Next, we check the light-like case. For example, turning on $\theta^{-12}$, 
where $x^{\pm}=\frac{1}{\sqrt{2}}(x^{5}\pm x^{0})$, gives the following 
open membrane data (Note that in this case (\ref{ommetric}) and 
(\ref{thetaabc}) cannot be used):
\begin{equation}\label{simpLL}
\tilde{G}_{\mu\nu}^{\sss {\rm OM}}=g_{\mu\nu}\ , \quad \Theta^{-12}_{\sss {\rm
 OM}}=\ell_{\rm p}^{3}\theta\ ,\quad \Theta^{-34}_{\sss {\rm OM}}=
\ell_{\rm p}^{3}\theta(g^{33}g^{44}g_{11}g_{22})^{1/2}\ .
\end{equation}
Note that if $g_{\mu\nu}$ is conformally flat $g^{33}g^{44}g_{11}g_{22}=1$ and
 $\Theta^{-12}_{\sss {\rm OM}}=\Theta^{-34}_{\sss {\rm OM}}$, see (\ref{M5LL})
for an example of this kind.

\subsection{Non-linear self-duality and the `tensor' equation of motion}
Next, we continue by showing that the three form obtained from (\ref{defM}) 
obeys the non-linear self-duality equation in the M5-brane directions. 
The non-linear self-duality equation on the M5-brane can be written as 
\cite{janpieter}:
\begin{eqnarray}\label{nle}
z^{-1}\tilde{G}_{\sss {\rm OM}}^{\mu\sigma}\tilde{A}_{\sigma}{}^{\nu\rho}&=&
\frac{1}{6}\frac{1}{\sqrt{-\tilde{g}_{(\mu\nu)}}}\epsilon^{\mu\nu\rho\mu'\nu'
\rho'}\tilde{A}_{\mu'\nu'\rho'}\ ,\\
z^{-1}&=&[K(1-\sqrt{1-K^{-2}})]^{1/3}\ ,
\end{eqnarray}
where $\epsilon^{012345}=-1$. We now check that (\ref{simpe}) obeys the 
non-linear self-duality equation (\ref{nle}). Inserting (\ref{simpe}) in the 
right hand side (RHS) and the LHS of (\ref{nle}), gives that RHS=LHS, i.e., 
the non-linear self-duality equation is satisfied. Repeating the 
calculation for a magnetic deformation also gives that (\ref{nle}) is 
satisfied. Note that for the non-linear self-duality to be satisfied 
$\omega=\sqrt{-g_{(\mu\nu)}}$ in (\ref{M5g}), i.e., the zero force condition 
has to be satisfied by the undeformed solution. That the non-linear self-dulity
equation is satisfied implies that electric and magnetic deformations give 
equivalent 
M5-M2 bound states\footnote{This is in agreement with results 
obtained in \cite{Berman}, for deformations of D3-branes (see section 5 and 
appendix B in \cite{Berman}). There it is shown that a magnetic deformation 
of a D3-brane is S-dual to an electric deformation of a D3-brane, if the 
undeformed D3-brane solution satisfies the zero force condition.}.

Inserting a light-like deformation also gives that (\ref{nle}) is satisfied. 
In this case there is a linear self-duality equation.

We note that showing that (\ref{nle}) is satisfied is equivalent to showing 
that the \emph{linear} self-duality equation (\ref{duality}) for 
$\Theta^{\mu\nu\rho}_{\sss {\rm OM}}$ is satisfied. Using (\ref{simpe1}), 
(\ref{simpm}) and (\ref{simpLL}), we obtain that the theta parameters, 
obtained in the last subsection, are linearly self-dual with respect to the 
open membrane metric. 

Next, we show that (\ref{defM}) satisfies $\ast \tilde{H}_{4}=\tilde{H}_{7}$ 
(with $\epsilon^{01\ldots 9,10}=-1$) under the 
restriction that all functions only depend on the transverse coordinates. Here
\begin{equation}\label{444}
\tilde{H}_{4}=d\tilde{A}_{3}\ ,\quad \tilde{H}_{7}=d\tilde{A}_{6}-\frac{1}{2}
\tilde{A}_{3}\wedge d\tilde{A}_{3}\ .
\end{equation}
For (\ref{defM}) we obtain 
\begin{equation}
\tilde{H}_{7}=dA_{6}-\tilde{A}_{3{\rm a}}\wedge d\tilde{A}_{3{\rm b}}-
(\tilde{A}_{3{\rm a}}+\tilde{A}_{3{\rm b}})\wedge d\gamma_{3}\ .
\end{equation}
For a one parameter deformation the relation 
$\ast \tilde{H}_{4}=\tilde{H}_{7}$ gives three independent relations
\begin{equation}\label{rel89}
(\ast_{\tilde{g}}) d\gamma_{3}=dA_{6}-\tilde{A}_{3{\rm a}}\wedge 
d\tilde{A}_{3{\rm b}}\ ,\quad (\ast_{\tilde{g}}) d\tilde{A}_{3{\rm a}}=
-\tilde{A}_{3{\rm b}}\wedge d\gamma_{3}\ ,\quad (\ast_{\tilde{g}}) 
d\tilde{A}_{3{\rm b}}=-\tilde{A}_{3{\rm a}}\wedge d\gamma_{3}\ . 
\end{equation}
We start by showing that the first relation in (\ref{rel89}) is correct. 
For an electric or a magnetic deformation we obtain that the left hand side 
(LHS) is
\begin{equation}\label{101}
(\ast_{\tilde{g}}) d\gamma_{3}=h^{-1}[(\ast_{g}) d\gamma_{3}]\ ,\quad 
h=1+\theta^{2}g_{(\alpha\beta)}\ ,
\end{equation}
while the right hand side (RHS) is 
\begin{equation}\label{102}
dA_{6}-\tilde{A}_{3{\rm a}}\wedge d\tilde{A}_{3{\rm b}}=h^{-1}d{A}_{6}\ ,
\end{equation}
where $\alpha,\beta=0,1,2$ for an electric deformation while 
$\alpha,\beta=3,4,5$ for a magnetic deformation. 
The result obtained in (\ref{101}) and (\ref{102}) implies that if the LHS 
should be equal to the RHS then 
$(\ast g) d\gamma_{3}=dA_{6}$. Since precisely this relation was demanded for 
the undeformed solution (\ref{M5g}), we have therefore shown that the first 
relation in (\ref{rel89}) is correct. For a light-like deformation the first 
relation in (\ref{rel89}) is trivially satisfied. Next, using that 
$(\ast_{\tilde{g}}) dA_{3(\rm b)}=h^{-1}[(\ast_{g}) dA_{3(\rm b)}]$, we 
 obtain that the third relation in (\ref{rel89}) is satisfied for all kinds 
of deformations. Further, since electric and magnetic deformations give 
equivalent solutions, the second relation must be correct due to the fact that 
the third relation is correct.

Next, we use that $\ast \tilde{H}_{4}=\tilde{H}_{7}$ in order to show that the 
eleven-dimensional supergravity `tensor' equation of motion
\begin{equation}\label{445}
d(\ast_{\tilde{g}}\tilde{H}_{4})=-\frac{1}{2}\tilde{H}_{4}\wedge 
\tilde{H}_{4}\ ,
\end{equation}
is satisfied. Using that $\ast \tilde{H}_{4}=\tilde{H}_{7}$ and (\ref{444}), 
we obtain that the LHS of (\ref{445}) is given by
\begin{equation}\label{446}
{\rm LHS}=-d\tilde{A}_{3{\rm a}}\wedge d\tilde{A}_{3{\rm b}}-
(d\tilde{A}_{3{\rm a}}+d\tilde{A}_{3{\rm b}})\wedge d\gamma_{3}\ ,
\end{equation}
while using that $\tilde{H}_{4}=d\tilde{A}_{3{\rm a}}+d\tilde{A}_{3{\rm b}}+
d\gamma_{3}$, we obtain that the RHS of (\ref{445}) is given by 
\begin{equation}\label{447}
{\rm RHS}=-d\tilde{A}_{3{\rm a}}\wedge d\tilde{A}_{3{\rm b}}-
(d\tilde{A}_{3{\rm a}}+d\tilde{A}_{3{\rm b}})\wedge d\gamma_{3}\ .
\end{equation}
Comparing (\ref{446}) and (\ref{447}) we find that (\ref{445}) is satisfied.

\section{Reduction to type IIA string theory}
In this section we are going to show that a double dimensional reduction of 
(\ref{defM}) for a one parameter deformation leads to the correct type IIA 
expression for a rank 2 NS-NS two form deformation of a D4-brane. Reducing 
longitudinal to the M5-brane but transverse to the deformation directions, 
leads to a new formula for one 
parameter three form RR deformations of D4-branes.

\subsection{Reduction to rank 2 $B$-field}

Next, we continue by showing that (\ref{defM}) gives (\ref{def}) (for $p=4$) 
with rank 2 $\theta_{\rm s}^{\mu\nu}$ under double dimensional reduction of 
$\theta^{\mu\nu\rho}$. We use the following relations between eleven and 
ten-dimensional fields (under the restriction $g_{\mu y}=0$):
\begin{eqnarray}\label{rel1}
\frac{g_{\sss MN}}{\ell_{\mathrm{p}}^2}&=&e^{-\frac{2\phi}{3}}\,
\frac{g^{(\rm s)}_{\sss MN}}{\alpha'}=\frac{g^{{\sss \mathrm{D}2}}_{\sss MN}}
{\alpha'}\,,\qquad\frac{g_{yy}}{\ell_{\mathrm{p}}^2}=
\frac{e^{\frac{4\phi}{3}}}{R^2}\ ,\nonumber\\
\frac{A_{3}}{\ell_{\mathrm{p}}^3}&=&\frac{C_{3}}{\alpha'^{\frac{3}{2}}}
+\frac{B_{2}}{\alpha'}\wedge \frac{dy}{R}\ ,\\
\frac{A_{6}}{\ell_{\mathrm{p}}^6}&=&\frac{B_{6}}{\alpha'^{3}}+
\Big(\frac{C_{5}}{\alpha'^{\frac{5}{2}}}+\frac{1}{2}\frac{C_{3}}
{\alpha'^{\frac{3}{2}}}\wedge \frac{B_{2}}{\alpha'}\Big)\wedge 
\frac{dy}{R}\ ,\nonumber
\end{eqnarray}
where $M,N=0,1,\ldots, 9$, $R$ is the radius of the compactified direction
 labeled by $y$ and $g_{\mu y}=0$. 
We also use the following standard parameter relations 
$\ell_{\rm{p}}^2=g^{\frac{2}{3}}\alpha'$ and $R=g\sqrt{\alpha'}$.

We start by setting $\theta_{\rm s}^{\alpha\beta}=\theta^{\alpha\beta y}$, 
where $\alpha,\beta=0,1, {\rm or}\, 3,4$, i.e., electric or magnetic, while 
$a,b$ are the other three directions and $y$ is the direction in which we 
reduce. This implies that 
\begin{equation}\label{thetar1}
\frac{1}{2}(\theta^{2})^{\alpha}{}_{\beta}=-(\theta_{\rm s}^{2})^{\alpha}{}
_{\beta}\ ,\quad (\theta)^{2}=-3(\theta_{\rm s})^{2}\ ,\quad 
\Big[\Big(1+\frac{1}{2}(\theta)^{2}\Big)^{-1}\Big]^{y}{}_{y}=[{\rm det}(1-
(\theta_{\rm s})^{2})]^{-1/2}\ .
\end{equation}
Further, we obtain that 
\begin{equation}\label{thetar2}
{\rm det}(1-(\theta_{\rm s})^{2})=\Big[{\rm det}\Big(1+\frac{1}{2}(\theta)^{2}
\Big)\Big]^{2/3}\ .
\end{equation}
Here we have used (\ref{thetar1}) and that 
\begin{equation}\label{r3}
{\rm det}\Big(1+\frac{1}{2}(\theta)^{2}\Big)=\Big(1+\frac{1}{6}(\theta)^{2}
\Big)^{3}\ ,\quad 
{\rm det}(1-(\theta_{\rm s})^{2})=\Big(1-\frac{1}{2}(\theta_{\rm s})^{2}
\Big)^{2}\ ,
\end{equation}
where we have used that $(\theta_{\rm s}^{2})_{\alpha\beta}=
(\theta_{\rm s})^{2}\frac{1}{2}g^{\rm (s)}_{\alpha\beta}$. 

We now use (\ref{rel1})-(\ref{r3}) in (\ref{defM}), which gives (\ref{def}) 
with $p=4$ and that $\theta_{\rm s}$ is rank 2 (electric or magnetic), i.e., a 
one parameter 
deformation. However, as has been shown in \cite{Delle}, the formula 
(\ref{def}) is valid also for a rank 4 deformation. The rank 4 case should 
be possible to obtain from a skew reduction of (\ref{defM}). 

The above calculations confirm that (\ref{defM}) under a double dimensional 
reduction of the three index theta, gives (\ref{def}) with $p=4$ and rank 2 
$\theta_{\rm s}$.  

\subsection{Reduction to one parameter RR three form deformation of a D4-brane}
In this subsection, we again reduce (\ref{defM}) to ten 
dimensions. But instead of reducing the three index theta to a two index theta
 we will reduce to a three index theta, i.e., reduce in a direction 
`transverse' to the deformation. This implies that we obtain a formula for 
deforming a D4-brane with a one parameter RR three form. Note that the 
undeformed solution is given in (\ref{ud}) with $p=4$. The reduction is 
straight forward and using (\ref{rel1}) we obtain (with $\mu,\nu=0,1,\ldots,4$)
\begin{eqnarray}\label{defD}
\tilde{g}^{\rm (s)}_{\mu\nu}&=&\Big[{\rm det}\Big(1+\frac{1}{2}(\theta)^{2}
\Big)\Big]^{1/6}g^{\rm (s)}_{\mu\rho}\Big[\Big(1+\frac{1}{2}(\theta)^{2}
\Big)^{-1}\Big]^{\rho}{}_{\nu}\ , \quad 
\tilde{g}^{\rm (s)}_{mn}=\Big[{\rm det}\Big(1+\frac{1}{2}(\theta)^{2}\Big)
\Big]^{1/6}g^{\rm (s)}_{mn}\ , \nonumber\\
g\tilde{C}_{3}&=&g\tilde{C}_{3{\rm a}}+\gamma_{3}\ ,\quad e^{2\tilde{\phi}}=
e^{2\phi}\Big[{\rm det}\Big(1+\frac{1}{2}(\theta)^{2}\Big)\Big]^{1/6}\ ,
\quad g^{2}\tilde{B}_{6}=\frac{1}{2}g\tilde{C}_{3{\rm a}}\wedge \gamma_{3}\ ,
\nonumber\\
g\tilde{C}_{5}&=&\omega dx^{0}\wedge \ldots \wedge dx^{4}-\tilde{B}_{2}\wedge 
\gamma_{3}\ ,\quad  \tilde{B}_{2}=-\omega\frac{1}{6}\theta^{\mu\nu\rho}
i_{\mu}i_{\nu}i_{\rho}dx^{0}\wedge\cdots \wedge dx^{4}\ ,\\
{\rm where}\nonumber \\
g\tilde{C}_{3{\rm a}}&=&\frac{1}{6}g\tilde{C}^{3{\rm a}}_{\mu\nu\rho}dx^{\mu}
\wedge dx^{\nu}\wedge dx^{\rho}\ ,\quad g\tilde{C}_{\mu\nu\rho}=-g^{2}
e^{-2\phi}g^{\rm (s)}_{\mu\rho'}g^{\rm (s)}_{\sigma\nu}
\theta^{\rho'\sigma\sigma'}g^{(\rm s)}_{\sigma'\lambda}\Big[\Big(1
+\frac{1}{2}(\theta)^{2}\Big)^{-1}\Big]^{\lambda}{}_{\rho}\ .\nonumber
\end{eqnarray}
Here $\theta^{\mu\nu\rho}$ is a dimensionless (one parameter) anti-symmetric 
tensor, and $(\theta^{2})^{\mu}{}_{\nu}$ is defined as follows:
\begin{equation}\label{theta33}
(\theta^{2})^{\mu}{}_{\nu}=g^{2}e^{-2\phi}\theta^{\mu\nu'\rho}g^{\rm (s)}_{
\nu'\sigma}g^{\rm (s)}_{\rho\sigma'}\theta^{\sigma\sigma'\lambda}g^{\rm (s)}_{
\lambda\nu}=g^{2}\theta^{\mu\nu'\rho}g^{\sss {\rm D2}}_{\nu'\sigma}
g^{\sss {\rm D2}}_{\rho\sigma'}\theta^{\sigma\sigma'\lambda}g^{\sss {\rm D2}}
_{\lambda\nu}\ .
\end{equation}
We note that since this formula has been obtained from a direct dimensional 
reduction of a one parameter formula, it is only valid for one parameter 
deformations. For example, deforming with both $\theta^{012}\neq 0$ and 
$\theta^{234}\neq 0$ is not possible using (\ref{defD}) since the non-zero 
RR one form would be missing.

The above obtained formula (\ref{defD}) can be used to deform D4-branes by 
turning on a non-zero RR three form. For the half-supersymmetric case we 
have checked that electric, magnetic and 
light-like deformations give the correct solutions corresponding to 
D4-D2, D4-F1 and D4-D2-F1-W bound states, respectively. These solutions in 
the particular form obtained here have been obtained before in 
\cite{soloper,solo2}, using other methods\footnote{In \cite{soloper} it was 
shown that deforming a half-supersymmetric D4-brane with, e.g., a rank 2 
magnetic $B$-field is equivalent to deforming a half-supersymmetric D4-brane 
with an electric RR three form.}. 
As an extra check that (\ref{defD}) is correct we have checked that 
T-dualizing in a direction parallel to the deformation followed by S-duality, 
gives (\ref{def}) with $p=3$, as expected.

Next, we derive that the above formula implies that the open D2-brane metric 
and generalized noncommutativity parameter are manifestly deformation 
independent under one parameter deformations. The open D2-brane metric 
and generalized noncommutativity parameter are given by \cite{openm}:
\begin{equation}\label{od2}
\tilde{G}^{{\sss \widetilde{{\rm OD}2}}}_{\mu\nu}= 
\Big[1+{1\over 6}\tilde{C}_3^2\Big]^{-{1\over
3}}\left(\tilde{g}^{{\sss {\rm D}2}}_{\mu\nu}+{1\over
2}(\tilde{C}_{3}^{2})_{\mu\nu}\right)\ ,
\end{equation}
\begin{equation}\label{OD2theta}
\Theta^{\mu_{1}\mu_{2}\mu_{3}}_{\widetilde{{\sss {\rm OD}2}}}=-(\alpha')^{
\frac{3}{2}}(1+\frac{1}{6}\tilde{C}_{3}^{2})^{\frac{1}{3}}
\tilde{g}^{\mu_{1}\nu_{1}}_{\sss {\rm D}2}\tilde{C}_{\nu_{1}\nu_{2}\nu_{3}}
\tilde{G}^{\nu_{2}\mu_{2}}_{{\sss {\rm OD}2}}\tilde{G}^{\nu_{3}\mu_{3}}_{{\sss
 {\rm OD}2}}\ ,
\end{equation}
where $\tilde{g}^{{\sss {\rm D}2}}_{\mu\nu}=e^{-{2\tilde{\phi}\over 3}}
\tilde{g}^{\rm (s)}_{\mu\nu}$
 is the closed (deformed) D$2$-brane metric, and 
\beq \label{C2}
(\tilde{C}^2_3)_{\mu\nu}=\tilde{g}^{\rho_1\sigma_1}_{{\sss {\rm D}2}}
\tilde{g}_{{\sss {\rm D}2}}^{\rho_2\sigma_2}\tilde{C}_{\rho_1\rho_2\mu}
\tilde{C}_{\sigma_1\sigma_2\nu}\ ,\quad \tilde{C}_{3}^2=\tilde{g}_{{\sss 
{\rm D}2}}^{\mu\nu}(\tilde{C}^2_{3})_{\mu\nu}\
.\eeq 
Using (\ref{defD}) and that $\theta^{012}=\theta$ \emph{or} $\theta^{234}=
\theta$, i.e., we have electric or magnetic deformation, gives
\begin{eqnarray}\label{eq1}
\tilde{g}^{{\sss {\rm D}2}}_{\alpha\beta}&=&\Big(1+\frac{1}{6}(\theta)^{2}
\Big)^{-2/3}g^{{\sss {\rm D}2}}_{\alpha\beta}\ ,\quad \tilde{g}^{{\sss 
{\rm D}2}}_{ab}=\Big(1+\frac{1}{6}(\theta)^{2}
\Big)^{1/3}g^{{\sss {\rm D}2}}_{ab}\ ,\nonumber\\
\frac{1}{2}(\tilde{C}^2_3)_{\alpha\beta}&=&\frac{1}{6}(\theta)^{2}\Big(1+
\frac{1}{6}(\theta)^{2}\Big)^{-2/3}g^{{\sss {\rm D}2}}_{\alpha\beta}\ ,\quad
\tilde{C}_{3}^2=(\theta)^{2}\ ,
\end{eqnarray}
where $\alpha,\beta=0,1,2, $ or $ 2,3,4$, while $a,b=3,4,$ or $0,1$, 
respectively, for electric and magnetic deformations. Next, using (\ref{eq1}) 
in (\ref{od2}) and (\ref{OD2theta}), gives the following \emph{deformation 
independent} open D2-brane metric and generalized noncommutativity parameter:
\begin{equation}
\tilde{G}_{\mu\nu}^{\sss {\rm D}2}=g^{{\sss {\rm D}2}}_{\mu\nu}\ , \quad 
\Theta^{\alpha\beta\gamma}_{\sss {\rm D}2}=g\alpha'^{3/2}\theta\epsilon^{
\alpha\beta\gamma}\ .
\end{equation}
Here $\epsilon^{012}=\epsilon^{234}=1$. Also for a light-like deformation 
we obtain deformation independence. It is important that we have obtained 
that the open D2-brane metric and generalized noncommutativity parameter are 
deformation independent, since this means that any one parameter 
deformation
of any kind of D4-brane solution\footnote{Of course under the restriction that
 the solution generating technique is valid.}, gives manifestly 
deformation independent open D2-brane metric and generalized noncommutativity
 parameter. In \cite{openm,solo2} this was shown for a one parameter 
deformation of the half-supersymmetric D4-brane. The result here is more 
general.

\section{Deformation of IIA/B NS5-branes with RR three or two forms}
In this section we derive formulas for deforming type IIA NS5-branes with 
one parameter RR three or two forms, respectively.
For the case with an NS5-brane with non-zero RR three form we also derive 
the open D2-brane coupling which, e.g., is relevant for the 
OD2-theory \cite{GMSS}.

\subsection{Deformation of NS5-branes with a RR two form}

In this subsection we show how a type IIB NS5-brane can be deformed by 
turning on a non-zero rank 2 RR two form. We start with the following 
undeformed type IIB NS5-brane solution   
\begin{eqnarray}\label{udNS52}
ds^{2}&=&g^{\rm (s)}_{\mu\nu}dx^{\mu}dx^{\nu}+g^{\rm (s)}_{mn}dx^{m}dx^{n}\ ,
\quad e^{2\phi}=g^{2}\hat{F}\ , 
\nonumber\\
g^{2}B_{6}&=&-\omega dx^{0}\wedge \cdots \wedge dx^{5}\quad B_{2}=\gamma_{2}\ ,
\end{eqnarray}
where $\hat{F}$ is some function, $\omega=g^{2}e^{-2\phi}\sqrt{-g^{\rm (s)}_{
(\mu\nu)}}$ due to the zero force condition, $g$ is the closed string coupling 
constant and 
$x^{\mu}$, $\mu=0,\ldots ,5$, are coordinates in the brane directions,  
while $x^{m}$, $m=6,\ldots ,9$, are coordinates in the transverse 
directions. Note that the two form $B=\gamma_{2}$ is dual to the six form 
$B_{6}$. For example, for a maximally supersymmetric NS5-brane $B\sim 
\epsilon_{2}$, where $d\epsilon_{2}$ is the volume form of the three-sphere. 
The above solution (\ref{udNS52}) is assumed to be T-dual to the undeformed 
type IIA solution given below in (\ref{udNS5}), and S-dual to (\ref{ud}) with
 $p=5$.

The easiest way to obtain the formula for a type IIB NS5-brane deformed by 
a rank 2 RR two form is to S-dualize the formula for a type IIB D5-brane 
deformed by a rank 2 $B$-field (see (\ref{def}) with $p=5$). We will use the 
following conventions for S-duality (assuming zero axion and only rank 2 
$B$-field)
\begin{eqnarray}\label{S2}
\frac{g^{\rm s}_{MN}}{\alpha'_{\rm s}}&=&e^{-\phi}\frac{g_{MN}}{\alpha'}\ ,
\quad e^{\phi_{\rm s}}=e^{-\phi}\ ,
\nonumber\\
\frac{B^{\rm s}_{2}}{\alpha'_{\rm s}}&=&\frac{C_{2}}{\alpha'}\ ,\quad 
\frac{C^{\rm s}_{2}}{\alpha'_{\rm s}}=-\frac{B_{2}}{\alpha'}\ , \quad 
\frac{C^{\rm s}_{4}}{(\alpha'_{\rm s})^{2}}=\frac{C_{4}}{(\alpha')^{2}}
+\frac{B_{2}}{\alpha'}\wedge \frac{C_{2}}{\alpha'}\ ,\\
\frac{B^{\rm s}_{6}}{(\alpha'_{\rm s})^{3}}&=&-\frac{C_{6}}{(\alpha')^{3}}-
\frac{1}{2}\frac{B_{2}}{\alpha'}\wedge \frac{C_{4}}{(\alpha')^{2}}\ ,\quad 
\frac{C^{\rm s}_{6}}{(\alpha'_{\rm s})^{3}}=\frac{B_{6}}{(\alpha')^{3}}-
\frac{1}{2}\frac{C_{2}}{\alpha'}\wedge \frac{C_{4}}{(\alpha')^{2}}\ ,\nonumber
\end{eqnarray}
where $\alpha'_{\rm s}=g\alpha'$ and the index s means the S-dualized 
quantity. 

Next, we are going to S-dualize (\ref{def}) with $p=5$ for a rank 2 NS-NS
deformation, where (\ref{def}) with $p=5$ restricted to rank 2 $B$-field is 
given by
\begin{eqnarray}\label{defD52}
\tilde{g}^{\rm (s)}_{\mu\nu}&=&g^{\rm (s)}_{\mu\rho}\Big[\Big(1-(\theta)^{2}
\Big)^{-1}\Big]^{\rho}{}_{\nu}\ , \quad 
\tilde{g}^{\rm (s)}_{mn}=g^{\rm (s)}_{mn}\ , \nonumber\\
g'\tilde{C}_{2}&=&\gamma_{2}\ ,\quad \tilde{B}_{2}=\tilde{B}_{2{\rm a}}\ ,
\quad e^{2\tilde{\phi}}=e^{2\phi'}\Big[{\rm det}\Big(1-(\theta)^{2}\Big)
\Big]^{-1/2}\ ,\nonumber\\
g'\tilde{C}_{4}&=&g'\tilde{C}_{4{\rm b}}-\tilde{B}_{2}\wedge \gamma_{2}\ ,
\quad g'^{2}\tilde{B}_{6}=\frac{1}{2}g'\tilde{C}_{4}\wedge \gamma_{2}\ ,
\nonumber\\
\tilde{C}_{6}&=&C_{6}-\tilde{B}_{2}\wedge \tilde{C}_{4{\rm b}}\ ,\quad 
g'\tilde{C}_{4{\rm b}}=\omega\frac{1}{2}\theta^{\mu\nu}i_{\mu}i_{\nu}
dx^{0}\wedge\cdots \wedge dx^{5}\ ,\\
{\rm where}\qquad \qquad& &\nonumber \\
\tilde{B}_{2{\rm a}}&=&\frac{1}{2}\tilde{B}^{2{\rm a}}_{\mu\nu}dx^{\mu}
\wedge dx^{\nu}\ ,\quad \tilde{B}_{\mu\nu}^{2{\rm a}}=
-g^{\rm (s)}_{\mu\rho}\theta^{\rho
\sigma}g^{(\rm s)}_{\sigma\lambda}\Big[\Big(1-(\theta)^{2}\Big)^{-1}
\Big]^{\lambda}{}_{\nu}\ ,\nonumber
\nonumber
\end{eqnarray}
and
\begin{equation}
(\theta^{2})^{\mu}{}_{\nu}=\theta^{\mu\nu'}g^{\rm (s)}_{\nu'\rho}
\theta^{\rho\sigma}g^{\rm (s)}_{\sigma\nu}\ .
\end{equation}
Note that the formulas for the metric, dilaton and NS-NS two form would 
not change if we considered rank 4 and rank 6 deformations. 
Continuing by S-dualizing 
(\ref{defD52}), using (\ref{S2}), we obtain the following 
formula for a type IIB NS5-brane deformed with a rank 2 RR two 
form (note that $B_{6}=-C_{6}$)
\begin{eqnarray}\label{defNS52}
\tilde{g}^{\rm (s)}_{\mu\nu}&=&\Big[{\rm det}\Big(1-(\theta)^{2}
\Big)\Big]^{1/4}g^{\rm (s)}_{\mu\rho}\Big[\Big(1-(\theta)^{2}
\Big)^{-1}\Big]^{\rho}{}_{\nu}\ , \quad 
\tilde{g}^{\rm (s)}_{mn}=\Big[{\rm det}\Big(1-(\theta)^{2}\Big)
\Big]^{1/4}g^{\rm (s)}_{mn}\ , \nonumber\\
\tilde{C}_{2}&=&\tilde{C}_{2{\rm a}}\ ,\quad \tilde{B}_{2}=\gamma_{2}\ ,\quad 
e^{2\tilde{\phi}}=e^{2\phi}\Big[{\rm det}\Big(1-(\theta)^{2}\Big)
\Big]^{1/2}\ ,\quad \tilde{C}_{6}=0\ ,\nonumber\\
\tilde{B}_{6}&=&B_{6}-\frac{1}{2}g\tilde{C}_{2}\wedge g\tilde{C}_{4}\ ,\quad 
g\tilde{C}_{4}=\omega\frac{1}{2}\theta^{\mu\nu}i_{\mu}i_{\nu}
dx^{0}\wedge\cdots \wedge dx^{5}\ ,\\
{\rm where}\qquad & &\nonumber \\
g\tilde{C}_{2{\rm a}}&=&\frac{1}{2}g\tilde{C}^{2{\rm a}}_{\mu\nu}dx^{\mu}
\wedge dx^{\nu}\ ,\quad g\tilde{C}^{2{\rm a}}_{\mu\nu}=g^{2}
e^{-2\phi}g^{\rm (s)}_{\mu\rho}\theta^{\rho\sigma}g^{(\rm s)}_{\sigma\lambda}
\Big[\Big(1-(\theta)^{2}\Big)^{-1}\Big]^{\lambda}{}_{\nu}\ ,\quad 
\nonumber
\end{eqnarray}
and
\begin{equation}
(\theta^{2})^{\mu}{}_{\nu}=g^{2}e^{-2\phi}\theta^
{\mu\nu'}g^{\rm (s)}_{\nu'\rho}\theta^{\rho\sigma}
g^{\rm (s)}_{\sigma\nu}=g^{2}\theta^{\mu\nu'}g^{\sss {\rm D}1}_{\nu'\rho}
\theta^{\rho\sigma}g^{\sss {\rm D}1}_{\sigma\nu}\ .
\end{equation}
This formula is valid for an NS5-brane deformed with a rank 2 RR two form, 
except for the expressions for the metric and RR two form which are valid 
also for rank 4 and 6 deformations, while the expression for the dilaton is  
valid for a rank 4 but not rank 6 deformation. 
    
Next, we show that the metric and RR two form, under a deformation of the  
NS5-brane, transform similarly to how the metric and the $B$-field transform 
 under a deformation of a D$p$-brane. We begin by introducing the following 
tensor
\begin{equation}\label{40}
F_{\mu\nu}=gg^{\sss {\rm D}1}_{\mu\nu}-gC_{\mu\nu}\ ,
\end{equation}
where $g^{\sss {\rm D}1}_{\mu\nu}=e^{-\phi}g^{\rm (s)}_{\mu\nu}$ and $g$ is 
the closed (fundamental) string coupling constant. 
If we now perform the same projective transformation for $F_{\mu\nu}$ as we 
did for the tensor $E_{\mu\nu}$ in section 2, we get (starting with 
$C_{\mu\nu}=0$)
\begin{equation}\label{r2}
\tilde{F}_{\mu\nu}=\Big(\frac{F}{\theta F+1}\Big)_{\mu\nu}=\Big(\frac{
gg^{\sss {\rm D}1}(1-\theta gg^{\sss {\rm D}1})}{(1+\theta gg^{\sss {\rm D}1})
(1-\theta gg^{\sss {\rm D}1}) }\Big)_{\mu\nu}\ .
\end{equation}
From (\ref{r2}) we easily obtain the closed D1-brane metric and RR 
two form $\tilde{C}_{2}$. If we compare with (\ref{defNS52}) we find that 
the RR two form is the same and computing the closed D1-brane metric, using 
the closed string metric and dilaton given in (\ref{defNS52}), we find the 
same answer as we obtained from (\ref{r2}). This implies that under a 
deformation with 
non-zero RR two form, the tensor $F_{\mu\nu}$ transforms by the projective 
transformation given in (\ref{r2}). The result in (\ref{r2}) is valid for 
 a rank $\leq 6$ RR two form. We also obtain from both (\ref{defNS52}) and 
(\ref{r2}) that the closed D1-brane metric only changes in the deformed 
`directions', although the closed fundamental string metric changes in all 
`directions'.

In section 2 the projective transformation of the tensor $E_{\mu\nu}$ was a 
consequence of combining $O(p+1,p+1)$ transformations, i.e., transformations in
 the T-duality group. Here instead the projective transformation of the tensor 
$F_{\mu\nu}$ can be seen as U-duality transformations. This is most easily 
seen by noticing that turning on, e.g., a rank 2 RR two form on the NS5-brane, 
can be seen from a fundamental string perspective to correspond to starting 
with an NS5-brane which is S-dualized to a D5-brane. Next, one deforms the 
D5-brane with a rank 2 $B$-field, using the $O(p+1,p+1)$ method, followed by 
S-duality. From a D1-brane perspective it is possible that the projective 
transformation of $F_{\mu\nu}$ can be seen to correspond to some kind of 
non-perturbative D1-brane `T-duality' followed by a gauge transformation and 
finally a new non-perturbative T-duality in the directions of the gauge 
transformation of $C_{2}$. It would be interesting to investigate this 
non-perturbative `T-duality' further, see section 6 for more comments. For 
related ideas concerning D1-brane `T-duality', see \cite{sen1}.
 
We end this subsection by showing how the open D1-brane metric and 
noncommutativity parameter can be related to the closed D1-brane metric and 
the RR two form, similarly to how open and closed fundamental string data 
were related in section 2. We start by defining a tensor $\xi^{\mu\nu}$ as 
follows
\begin{equation}
\xi^{\mu\nu}=g^{-1}G_{\sss {\rm OD}1}^{\mu\nu}-
\frac{\Theta^{\mu\nu}_{\sss {\rm OD1}}}{g\alpha'}\ .
\end{equation}
Similarly to section 2, we obtain that $F$ and $\xi$ are related through
\begin{equation}\label{relD1}
\xi^{\mu\nu}=(F^{-1})^{\mu\nu}\ .
\end{equation}
This implies that the open D1-brane metric and noncommutativity parameter
 are manifestly deformation independent under deformations with an RR two 
form of an NS5-brane. It is interesting that we have found that several 
of the properties of how open and closed strings transform are the same 
for D1-branes and fundamental strings. Further, considering the 
conjectured SL$(2,\mathbf{Z})$ symmetry of type IIB string theory we expect it
 to be possible to generalize (\ref{relos}) and (\ref{relD1}) to (p,q) 
strings. We also expect it to be possible to generalize the $O(p+1,p+1)$ 
method (\ref{def}) not only to deformations of NS5-branes with a RR two 
form (\ref{defNS52}) but to deform (p,q) 5-branes with a `(p,q)' theta 
deformation (i.e., turning on a combination of RR and NS-NS two forms). 
This can be obtained by performing an SL$(2,\mathbf{Z})$ transformation of 
(\ref{def}) with $p=5$. This would be a different 
approach to deforming (p,q) 5-branes than in \cite{martin,mikkel}, where 
instead the equations of motion were explicitly solved.

\subsection{Deformation of NS5-branes with a RR three form}
In this subsection we are going to generalize the results for deforming 
the type IIB NS5-brane to deformations of type IIA NS5-branes with a one 
parameter RR 
three form. We start with the following undeformed solution 
\begin{eqnarray}\label{udNS5}
ds^{2}&=&g^{\rm (s)}_{\mu\nu}dx^{\mu}dx^{\nu}+g^{\rm (s)}_{mn}dx^{m}dx^{n}\ ,
\quad e^{2\phi}=g^{2}F'\ , 
\nonumber\\
g^{2}B_{6}&=&-\omega dx^{0}\wedge \cdots \wedge dx^{5}\quad B_{2}=\gamma_{2}\ .
\end{eqnarray}

Next, generalizing the results for the type IIB NS5-brane is  
straight forward and gives
\begin{eqnarray}\label{defNS5}
\tilde{g}^{\rm (s)}_{\mu\nu}&=&\Big[{\rm det}\Big(1+\frac{1}{2}(\theta)^{2}
\Big)\Big]^{1/6}g^{\rm (s)}_{\mu\rho}\Big[\Big(1+\frac{1}{2}(\theta)^{2}
\Big)^{-1}\Big]^{\rho}{}_{\nu}\ , \quad 
\tilde{g}^{\rm (s)}_{mn}=\Big[{\rm det}\Big(1+\frac{1}{2}(\theta)^{2}\Big)
\Big]^{1/6}g^{\rm (s)}_{mn}\ , \nonumber\\
\tilde{C}_{3}&=&\tilde{C}_{3{\rm a}}+\tilde{C}_{3{\rm b}}\ ,\quad 
\tilde{B}_{2}=\gamma_{2}\ ,\quad e^{2\tilde{\phi}}=e^{2\phi}\Big[{\rm det}
\Big(1+\frac{1}{2}(\theta)^{2}\Big)\Big]^{1/6}\ ,\nonumber\\
\tilde{B}_{6}&=&B_{6}+\frac{1}{2}\tilde{C}_{3{\rm a}}\wedge 
\tilde{C}_{3{\rm b}}\ ,\quad \tilde{C}_{5}=0\ ,\\
{\rm where}\nonumber \\
g\tilde{C}_{3{\rm a}}&=&\frac{1}{6}g\tilde{C}^{3{\rm a}}_{\mu\nu\rho}dx^{\mu}
\wedge dx^{\nu}\wedge dx^{\rho}\ ,\quad g\tilde{C}^{3{\rm a}}_{\mu\nu\rho}=
-g^{2}e^{-2\phi}g^{\rm (s)}_{\mu\rho'}g^{\rm (s)}_{\sigma\nu}
\theta^{\rho'\sigma\sigma'}g^{(\rm s)}_{\sigma'\lambda}\Big[\Big(1
+\frac{1}{2}(\theta)^{2}\Big)^{-1}\Big]^{\lambda}{}_{\rho}\ ,\quad 
\nonumber\\
g\tilde{C}_{3{\rm b}}&=&\omega\frac{1}{6}\theta^{\mu\nu\rho}i_{\mu}i_{\nu}
i_{\rho}dx^{0}\wedge\cdots \wedge dx^{5}\ .\nonumber
\end{eqnarray}

Here $\theta^{\mu\nu\rho}$ is a dimensionless (one parameter) anti-symmetric 
tensor, and $(\theta^{2})^{\mu}{}_{\nu}$ is defined as in (\ref{theta33}).

The above obtained formula (\ref{defNS5}) can be used to deform NS5-branes by 
turning on a non-zero RR three form. Below we show that if the above 
formula (\ref{defNS5}) first is T-dualized in a direction parallel to the 
deformed direction followed by S-duality, we obtain (\ref{def}) with $p=2$ and
 rank 2 $B$-field. This shows that (\ref{defNS5}) is correct. 
Further, for the half-supersymmetric case we have checked that electric, 
magnetic and 
light-like deformations, give the correct solutions corresponding to 
NS5-D2 and NS5-D2-D2-W bound states, respectively. These solutions in 
the particular form obtained here have been obtained before in 
\cite{soloper,solo2}, using other methods. We note that similarly to the 
M5-brane case, electric and magnetic deformations give equivalent NS5-D2 
solutions \cite{soloper}. 

Also similarly to the eleven-dimensional case, the three form
$\tilde{C}_{3}$ obeys a non-linear self-duality condition in the NS5-brane 
directions. Further, one can show that as in the M5-brane case, 
electric and light-like deformations give deformation independent open 
D2-brane metric and generalized noncommutativity parameter\footnote{The 
open D2-brane metric and generalized noncommutativity parameter are 
essentially given by the same expressions as the open M2-brane data 
(\ref{OMdata}), see \cite{openm}.}. 

Next, to show that (\ref{defNS5}) is correct, we are going to show 
that (\ref{defNS5}) is T-dual to (\ref{defNS52}). For T-duality we use the 
same conventions as was used in \cite{solo2}.
T-dualizing (\ref{defNS5}) in one of the 
deformed directions and identifying $\theta^{\mu\nu}=-\theta^{\mu\nu y}$, where
 $y$ is the T-dualized direction, and using that  
\begin{equation}
\frac{1}{2}(\theta^{2})^{\alpha}_{}{\sss {\beta}}\rightarrow 
-(\theta^{2})^{\alpha}_{}{\sss {\beta}}\ ,\quad {\rm det}\Big(1+\frac{1}{2}
(\theta)^{2}\Big)\rightarrow [{\rm det}(1-(\theta)^{2})]^{3/2}\ ,
\end{equation}
we obtain (\ref{defNS5}), as expected.

The solution generating technique (\ref{defNS5}) is related to (\ref{defM}) by 
a lift. We note that if (\ref{defNS5}) is lifted to eleven dimensions we 
obtain a slightly different version than (\ref{defM}) (there are minus signs 
which differ in a few places). However, the obtained formula is equivalent 
to (\ref{defM}). The reason for this difference lies in how (\ref{defNS5}) 
has been obtained. As is clear from above, (\ref{defNS5}) has been obtained
 by first S-dualizing the $O(p+1,p+1)$ method for $p=5$ and rank 2 $B$-field, 
followed by T-duality. We next give an example which illuminates and explains 
why there is a sign difference when lifting (\ref{defNS5}) to eleven 
dimensions, compared to (\ref{defM}). 

Start with an M5-M2 solution which is smeared in 
the $x^{6}$ direction. The M5-brane is in the 1-5 directions and the M2-brane 
in the 1 and 5 directions. The M-theory three form is given by 
\begin{equation}\label{321}
A_{3}=+Adx^{0}\wedge dx^{1}\wedge dx^{5}-Bdx^{2}\wedge dx^{3}\wedge dx^{4}+
\gamma_{2}\wedge dx^{6}\ ,
\end{equation}
while $\epsilon^{0123456789,10}=-1$. Note that in this example the exact form
 of the functions $A$ and $B$ is not interesting. 
Next, we compactify on a two torus in the 5 and 6 directions. This implies 
that we have the following parameters: $\ell_{p}$, $R_{5}$ and $R_{6}$, where 
$R_{5}$ and $R_{6}$ are the radius in the 5 and 6 directions, respectively. 
We will now reduce in the 5 direction to type IIA and perform the following 
series of dualities: T-duality in the 6 direction, S-duality, T-duality in the
 6 direction and finally we lift to M-theory (in the 5 direction). After 
performing these dualities we obtain an M5-M2 solution with the following 
three form:
\begin{equation}\label{322}
A_{3}=-Adx^{0}\wedge dx^{1}\wedge dx^{6}-Bdx^{2}\wedge dx^{3}\wedge dx^{4}+
\gamma_{2}\wedge dx^{5}\ ,
\end{equation}
while $\tilde{\ell_{p}}=\ell_{p}$, $\tilde{R}_{5}=R_{6}$ and
$\tilde{R}_{6}=R_{5}$. This means that we have obtained an M5-M2 solution 
smeared in the 5 direction and where the M5-brane is in the 1-4 and 6 
directions, while the M2-brane is in the 1 and 6 directions. What we have 
done here is a U-duality transformation which have switched place between 
the 5 and 6 directions. If we now let $5\rightarrow 6$ and $6\rightarrow 5$, 
(\ref{322}) becomes 
\begin{equation}\label{323}
A_{3}=-Adx^{0}\wedge dx^{1}\wedge dx^{5}-Bdx^{2}\wedge dx^{3}\wedge dx^{4}+
\gamma_{2}\wedge dx^{6}\ .
\end{equation}
However, and most importantly, epsilon changes to 
$\epsilon^{0123456789,10}=+1$. The conclusion is that the sign change 
in one component of $A_{3}$ is compensated by the sign change in epsilon. 
This implies that if (\ref{321}) solves the eleven-dimensional equations of 
motion, then so does (\ref{323}), as long as also epsilon is changed. 
This explains why there is a sign difference when lifting (\ref{defNS5}) to 
eleven dimensions, compared to (\ref{defM}).
  
We will end this subsection by deriving an open D2-brane coupling 
$G^{2}_{\sss {\rm OD}2}$ for an open D2-brane ending on an NS5-brane, using 
(\ref{defNS5}) and deformation independence. This open D2-brane coupling was 
first postulated in \cite{GMSS} (see also \cite{soloper}) as the relevant 
coupling for the OD2-theory. Here we derive a covariant expression for 
this open D2-brane coupling which in the OD2 limit gives the coupling 
introduced in \cite{GMSS,soloper}. We start with the following ansatz 
for the open D2-brane coupling
\begin{equation}\label{ansatz2}
G^{2}_{\sss {\rm OD}2}=e^{\tilde{\phi}}R(K)\ ,
\end{equation}
where $R(K)$ is a function of 
\begin{equation}
K=\sqrt{1+\frac{1}{24}\tilde{C}^{2}}\ ,\quad \tilde{C}^2=\tilde{g}^{\mu_{1}
\nu_{1}}\tilde{g}^{\mu_{2}\nu_{2}}\tilde{g}^{\mu_{3}\nu_{3}}
\tilde{C}_{\mu_{1}\mu_{2}\mu_{3}}\tilde{C}_{\nu_{1}\nu_{2}\nu_{3}}\ .
\end{equation}
Next, we are going to assume that the open D2-brane coupling is deformation 
independent for electric and light-like deformations (similar to the 
 open D2-brane metric and generalized noncommutativity parameter). 
This implies that  
\begin{equation}\label{ansatz3}
G^{2}_{\sss {\rm OD}2}=e^{\tilde{\phi}}R(K)=e^{\phi}\ ,
\end{equation}
where $e^{\phi}$ is the undeformed dilaton. For a light-like deformation 
this is trivially satisfied for any $R(K)$. However, for an electric 
deformation (i.e., $\theta^{012}=\theta$, $\alpha,\beta=0,1,2,$) we obtain, 
using (\ref{defNS5}) and (\ref{ansatz3}), that
\begin{equation}
R(K)=h^{-1/4}\ ,\quad h=1+\theta^{2}g^{2}g^{\sss {\rm D}2}_{(\alpha\beta)}\ ,
\end{equation}
where $g^{\sss {\rm D}2}_{(\alpha\beta)}={\rm det}g^{\sss {\rm D}2}_{\alpha
\beta}$, and $g^{\sss {\rm D}2}_{\alpha\beta}=e^{-\frac{2}{3}\phi}
g^{\rm (s)}_{\alpha\beta}$ is the undeformed closed D2-brane metric. Using 
that $0<h<1$ we get that $R(K)=[K(1-\sqrt{1-K^{-2}})]^{-1/2}$ (this expression 
can also be rewritten as $R(K)=[K(1+\sqrt{1-K^{-2}})]^{1/2}$). This leads 
to the following open D2-brane coupling
\begin{equation}\label{OD2c}
G^{2}_{\sss {\rm OD}2}=e^{\tilde{\phi}}[K(1-\sqrt{1-K^{-2}})]^{-1/2}\ .
\end{equation}
As a check of (\ref{OD2c}) we have inserted the OD2-limit \cite{GMSS,soloper} 
into (\ref{OD2c}) and found that it gives exactly the OD2 coupling defined in 
\cite{GMSS,soloper}. It would be very interesting to derive this expression 
from a microscopic formulation of an open D2-brane ending on an NS5-brane.
For a discussion of covariant expressions for open D$p$-brane couplings 
($p\neq 2$), see \cite{mikkel6}.

\section{$p$-branes and SL$(2,\mathbf{R})$} 
\subsection{M-theory and SL$(2,\mathbf{R})$}
In this subsection we will argue that the metric and three form in the 
deformed `directions', obtained from the solution generating technique 
(\ref{defM}), for a magnetic deformation, can be seen to emerge from certain 
SL$(2,\mathbf{R})$ transformations of the three-torus K$\ddot{{\rm a}}$hler 
structure parameter\footnote{For related issues see \cite{sen1,aha1}.} 
(we choose e.g., $\theta^{345}=\theta$), 
\begin{equation}
E=A_{345}+i\sqrt{{\rm det}g_{ab}}\ ,
\end{equation}
where $A_{345}=0$, in the initial solution. As has been explained in section 
3, when deforming an M5-brane one starts by compactifying the M5-brane on a 
three-torus etc, see section 3. This suggests that in order to obtain a new 
deformed solution we should perform the following SL$(2,\mathbf{R})$ 
transformation of the K$\ddot{{\rm a}}$hler structure parameter:
\begin{equation}\label{987}
\tilde{E}=\frac{E}{-\theta E+1}=\frac{-\theta{\rm det}g_{ab}+
i\sqrt{{\rm det}g_{ab}}}{1+\theta^{2}{\rm det}g_{ab}}\ .
\end{equation}
This implies that 
\begin{eqnarray}\label{sl2}
\tilde{g}_{ab}&=&\frac{g_{ab}}{(1+\theta^{2}{\rm det}g_{ab})^{2/3}}\ ,
\nonumber\\
\tilde{A}_{345}&=&\frac{-\theta{\rm det}g_{ab}}{1+\theta^{2}{\rm det}
g_{ab}}\ .
\end{eqnarray}
Note that we can only expect the transformation (\ref{987}) to be valid for a 
magnetic deformation, since electric and light-like deformations involve 
the time direction. However, the deformed metric and three form 
can be obtained from (\ref{sl2}) also for electric deformations 
due to analytic continuation from the magnetic case. Comparing (\ref{sl2}) 
with the metric and three form (in the deformed
`directions') in (\ref{defM}) we obtain that they are identical for 
magnetic (and electric) deformations. For the magnetic deformations the above 
means that the metric and three form in the deformed `directions' transform 
`together' under the above SL$(2,\mathbf{R})$ transformation, when deforming 
the M5-brane. 

The above SL$(2,\mathbf{R})$ transformation is easily seen to be given by
\begin{equation}\label{trans1}
\tilde{E}=({\rm S}^{-1}{\rm T}{\rm S})E\ ,
\end{equation}
where 
\begin{equation}
{\rm T}=\pmatrix{1&\theta\cr 0&1}\ ,\quad {\rm S}=\pmatrix{0&1\cr -1&0}\ .
\end{equation}
Note that a general SL$(2,\mathbf{R})$ transformation, i.e., $\tau\rightarrow
 \frac{a\tau+b}{c\tau+d}$, is given by the following matrix
\begin{equation}
\pmatrix{a&b\cr c&d}\ ,
\end{equation}
where $ad-bc=1$.

The transformation given in (\ref{trans1}) implies that we first invert the 
volume of the three-torus (i.e., we use the $S$ transformation on the 
K$\ddot{{\rm a}}$hler structure parameter\footnote{In \cite{sen1} this 
transformation was named `T-duality' for M2-branes, because it exchanges 
the Kaluza-Klein modes with the wrapping modes of the M2-brane.})  
followed by a 
gauge transformation and finally, we invert the volume of the torus again. 
Since the SL$(2,\mathbf{R})$ transformations we performed here gave the 
same result as in (\ref{defM}), for the magnetic case, we have yet another 
non trivial test of (\ref{defM}). Also, viewing the deformations as 
certain SL$(2,\mathbf{R})$ transformations makes the deformation procedure 
more transparent as U-duality transformations. We note that SL$(2,\mathbf{R})$
is part of the U-duality group SL$(3,\mathbf{R})\times$ SL$(2,\mathbf{R})$
 for eleven-dimensional supergravity compactified on a three-torus 
\cite{hull}. In the full M-theory the U-duality group is expected to be 
SL$(3,\mathbf{Z})\times$ SL$(2,\mathbf{Z})$ \cite{hull}.

It is easy to see that (\ref{trans1}) is the only non-trivial deformation 
which is 
possible to perform if there has to be an equal number of S and ${\rm S}^{-1}$.
 The reason for the equal number of S and ${\rm S}^{-1}$ is because S can be 
viewed as `inverting' the  three torus, while 
 ${\rm S}^{-1}$ is just the inverse of this transformation. Therefore, 
an equal number of S and ${\rm S}^{-1}$ implies that starting with an 
M5-brane we end with a (deformed) M5-brane. Note that an odd total number of S 
and ${\rm S}^{-1}$ transformations would imply that we end with an M2-brane 
solution.

The next step towards a further understanding of (\ref{defM}), would be 
to obtain exactly how the tensors ($g_{\mu\nu}$ and $A_{\mu\nu\rho}$) 
transform and not 
only the complex scalar ($E$), since this should lead to  
a better understanding about also the electric and light-like cases. This is 
in contrast to the string case (see 
section 2 and 5) where we know how the \emph{tensors} $E_{\mu\nu}$ and 
$F_{\mu\nu}$ transform. For example, is it (in the M-theory case) possible to 
form a three index tensor out of 
the metric and three form which transforms in some `nice' way? This three 
index tensor should be possible to obtain since we know from (\ref{defM})  
the end result of this transformation (i.e., from (\ref{defM}) we obtain the 
end result for $g_{\mu\nu}$ and $A_{\mu\nu\rho}$). This new three 
index tensor should also be the generalization of the two index tensor
$E_{\mu\nu}=g_{\mu\nu}+B_{\mu\nu}$ to eleven dimensions. 
So far, we have unfortunately not been able to obtain this three index 
tensor. We plan to discuss this further in a future paper \cite{solo3}.
\subsection{D-branes and SL$(2,\mathbf{R})$}
In this subsection we show how the metric and $(p+1)$-form $C$ ($p=0,1,2,3,4$)
for a D($p+2$)-brane or NS5-brane, transform in the deformation directions, 
under a magnetic deformation with a constant anti-symmetric 
($p+1$)-form $\theta^{p+1}$. Similarly to the M-theory case in the last 
subsection we start by giving the K$\ddot{{\rm a}}$hler structure 
parameter for a NS5-brane or a D($p+2$)-brane compactified on a 
($p+1$)-torus ($a,b=1,2,\ldots,p+1$)
\begin{equation}
E=gC_{12\ldots (p+1)}+ig\sqrt{{\rm det}g^{\sss {\rm D}p}_{ab}}\ ,
\end{equation}
where $g$ is the closed string coupling constant and $C_{12\ldots (p+1)}=0$, 
in the initial solution. To obtain a new 
deformed solution we perform the same SL$(2,\mathbf{R})$ transformation 
 (\ref{trans1}) as we did in the M-theory case. This gives the following 
result
\begin{eqnarray}\label{sl2D}
\tilde{g}^{\sss {\rm D}p}_{ab}&=&\frac{g^{\sss {\rm D}p}_{ab}}{(1+\theta^{2}
g^{2}{\rm det}g^{\sss {\rm D}p}_{ab})^{\frac{2}{p+1}}}\ ,
\nonumber\\
g\tilde{C}_{12\ldots (p+1)}&=&\frac{-\theta g^{2}{\rm det}g_{ab}}{1+
\theta^{2}g^{2}{\rm det}g_{ab}}\ .
\end{eqnarray}
For $p=2$ we compare (\ref{sl2D}) with the metric and three form (in the theta
 `directions') in (\ref{defNS5}) and (\ref{defD}), for a magnetic deformation,
 and obtain that they are 
identical. For $p=1$ we also obtain perfect agreement if we let 
$\theta\rightarrow -\theta$ in (\ref{sl2D}). 

We note that, as expected, for both electric and magnetic deformations, 
the open D$p$-brane metric is deformation independent and the generalized 
noncommutativity parameter is shifted by a constant $\sim \theta$.
It is interesting to note that since the open D$p$-brane metric is 
deformation independent and the generalized noncommutativity parameter is 
shifted by a constant $\sim \theta$, when $E$ transforms according to 
(\ref{trans1}), the transformation of the open D$p$-brane
 data (in the deformed direction) can be written as follows
\begin{equation}\label{open1}
\tau\rightarrow {\rm T}\tau\ .
\end{equation}
Here 
\begin{equation}
\tau=ig^{-1}\sqrt{{\rm det}(G)^{-1}_{\sss {\widetilde{{\rm OD}p}}}}+
\frac{\Theta_{{\sss {\widetilde{{\rm OD}p}}}}}{g\alpha'^{\sss 
{\frac{p+1}{2}}}}\ ,
\end{equation}
while ${\rm det}(G)^{-1}_{\sss {\widetilde{{\rm OD}p}}}$ is the determinant of
 the inverse of the open D$p$-brane metric and 
$\Theta_{{\sss {\widetilde{{\rm OD}p}}}}$ is the open D$p$-brane 
generalized noncommutativity parameter. Note that before the deformation 
$\tau=({\rm S})E$, i.e., (\ref{open1}) implies 
that $\tau\rightarrow \tau=({\rm T}{\rm S})E$, since \mbox{$\Theta_{{\sss 
{\widetilde{{\rm OD}p}}}}=\theta g\alpha'^{\sss {\frac{p+1}{2}}}$}.
This means that when the closed 
D$p$-brane data (i.e., $C_{\rm p+1}$ and $g^{\sss {\rm D}p}_{ab}$ in the 
deformed directions) transform as in (\ref{trans1}), the open D$p$-brane 
data (metric and theta parameter in the deformed directions) have the 
simple transformation (\ref{open1}). Further, after the deformation the new 
$\tau$ can be seen to be 
\begin{equation}\label{rel8}
\tilde{\tau}={\rm S}\tilde{E}\ .
\end{equation}
Comparing (\ref{rel8}) with the open D$p$-brane data obtained in \cite{openm}, 
we find that (\ref{rel8}) gives the correct  
open D$p$-brane metric and generalized noncommutativity parameter 
(in the deformed directions). 

It is interesting that the relation between the open D$p$-brane data and 
`closed' 
D$p$-brane data in (\ref{rel8}) is very similar to how the open and closed 
string (D1-brane) data are related in (\ref{relos}) (and (\ref{relD1})). We do
 not yet know how relevant 
this result is. However, it is probably yet another indication that the 
open D$p$-brane metrics and generalized noncommutativity parameters, obtained 
in \cite{openm}, are correct.

\section{Conclusions}

In this paper we have obtained an M-theory solution generating technique, which
 can be used to deform an M5-brane with a non-zero three form $A$. To check 
that this solution generating technique gives the correct results we have 
performed several tests, both directly in eleven dimensions (see 
section 3) and by reduction to ten dimensions and comparing with known 
results (see section 4). For example, we showed that the eleven-dimensional 
`tensor' equation is satisfied and that (\ref{defM}) gives the correct 
half-supersymmetric bound state solutions M5-M2 and M5-M2-M2-MW. Together, the
 tests that we have performed in ten and eleven dimensions,
strongly indicate that we have obtained the correct solution generating 
technique. Note, however, that we have not rigorously 
proved that (\ref{defM}) is correct. A rigorous proof would be to also show 
that the eleven-dimensional `Einsteins' equation of motion is satisfied by 
(\ref{defM}). This seems difficult to show, because  
`Einsteins' equation of motion is second order in derivatives of the deformed 
metric. This is an important difference compared to the eleven-dimensional
`tensor' equation, which avoids explicit derivatives on the deformed metric. 

In section 6 we have shown that for a magnetic deformation, it is possible 
to view the deformation as certain SL$(2,\mathbf{R})$ transformations of the 
 K$\ddot{{\rm a}}$hler structure parameter for the three-torus, on which the 
M5-brane has been compactified. This is an important result since it shows 
that (as expected) deforming an M5-brane corresponds to performing the 
appropriate U-duality transformations\footnote{Note that SL$(2,\mathbf{R})$
is part of the U-duality group SL$(3,\mathbf{R})\times$ SL$(2,\mathbf{R})$
 for eleven-dimensional supergravity compactified on a three-torus 
\cite{hull}.}, i.e., it is correct to view the 
deformation procedure as first performing (an M-theory version of) `T-duality'
 in three directions, followed by a gauge transformation, and finishing with 
`T-duality' in the same three directions as before. 

We have also in this paper obtained solution generating techniques for 
deforming the type IIA/B NS5-branes with a one parameter RR three or two form,
 respectively. These solution generating techniques were shown to generate 
the expected results. Further, in the type IIA case we used the newly obtained 
solution generating technique and deformation independence to derive a 
covariant expression for an open D2-brane coupling, relevant for OD2-theory. 
It would be very interesting if this result could be derived from a 
microscopic formulation of an open D2-brane ending on an NS5-brane. 

In a future paper \cite{solo3} we plan to expand on the results in section 6 
to tensor relations for $A_{\mu\nu\rho}$ and $g_{\mu\nu}$, in order to obtain 
a better understanding of M-theory `T-duality'. For example, it would be 
interesting to see if it is possible to derive some kind of `generalization' 
of the relation $E_{\mu\nu}=g_{\mu\nu}+B_{\mu\nu}$ in string theory to 
M-theory? Our two main motivations for a further study of M-theory 
`T-duality', is to obtain more information about the open membrane metric and 
generalized noncommutativity parameter, 
which are deformation independent under certain M-theory `T-duality' 
transformations (i.e., under the combination `T-duality' + gauge 
transformation + `T-duality', see \cite{openm} and section 3 and 6 for more 
details), and to also be able to derive (i.e., to rigorously prove) the 
solution generating technique given in (\ref{defM}), from a microscopic 
formulation of M-theory `T-duality'.

\vspace{0.4cm}
\Large \textbf{Acknowledgments}
\vspace{0.2cm}
\normalsize

We are grateful to R. Argurio, P. Arvidsson, M. Cederwall, U. Gran, 
M. Nielsen, B.E.W. Nilsson and P. Sundell for valuable discussions and 
comments.

\end{document}